\documentclass{jfm}

\usepackage{amsmath,amssymb,mathtools,bm}
\usepackage{multirow}
\usepackage[hidelinks]{hyperref}
\usepackage[capitalise,nameinlink]{cleveref}
\usepackage{graphicx}
\usepackage{overpic}
\usepackage{tabularx}
\usepackage[normalem]{ulem}
\newcolumntype{Y}{X}
\usepackage{pdfpages}
\usepackage{tabto}
\usepackage{amsmath}
\usepackage{nicematrix}

\usepackage{hyperref}

\usepackage{array}
\newcolumntype{L}[1]{>{\raggedright\let\newline\\\arraybackslash\hspace{0pt}}m{#1}}
\newcolumntype{C}[1]{>{\centering\let\newline\\\arraybackslash\hspace{0pt}}m{#1}}
\newcolumntype{R}[1]{>{\raggedleft\let\newline\\\arraybackslash\hspace{0pt}}m{#1}}

\usepackage{relsize,exscale}
\usepackage[skip=10pt plus1pt, indent=10pt]{parskip}
 
\usepackage{my_macros}
\usepackage{natbib}
\title{Modelling the impact of synovial fluid elasticity on tangential stress}
\shorttitle{Modelling synovial fluid elasticity and tangential stress}
\shortauthor{Gaffney, Brown and Whiteley}

\author{Eamonn A. Gaffney\aff{1*}, Cameron P. Brown\aff{2}, Jonathan P. Whiteley\aff{3}}

\affiliation{\aff{1} Wolfson Centre for Mathematical Biology, Mathematical Institute, University of Oxford, Oxford,  UK
\aff{2}QUT Medical Engineering Research Facility;  MMPE, FoE, Queensland University of Technology, Australia
\aff{3}Department of Computer Science, University of Oxford, Oxford,  UK
\aff{*} Corresponding author, email: gaffney@maths.ox.ac.uk}
 
\usepackage{my_macros}
\usepackage{natbib}

\renewcommand{\vec}[1]{\bm{#1}}
\newcommand{\tensor}[1]{\bm{#1}}

\newcommand{\e}{\mathrm{e}}

\newcommand \bl{\color{black}} \newcommand \mg{\color{black}}

\begin{document}

\maketitle 

\date{\today}
 
\begin{abstract}
The rheological properties of synovial fluid have been observed to substantially impact its lubricating behaviour. While numerous studies have illustrated the importance of its shear-dependent viscosity,  the impact of synovial fluid elasticity  for oscillatory joint motion is far less characterised. Hence we consider how elasticity impacts the tangential stress, and thus friction, exerted on confining surfaces in  rheological models of synovial fluid on the length and time scales of oscillatory joint motion, though in a simplified setting rather than considering the full complexity of a joint.  {\mg Minor changes in the elastic constitutive equation from a canonical upper convected Oldroyd-B model   lead  to fundamentally different scalings and predictions for tangential stress compared to either  a Newtonian fluid and an Oldroyd-B fluid. In particular, 
 polymer elasticity within physiological fluids is predicted to have a profound effect on friction both within the oscillating joint and more generally, in turn  suggesting further examination of synovial fluid elasticity in experimental and modelling studies. }

\vspace*{0.2cm}\noindent

\end{abstract}
\section{Introduction}

\begin{figure}
 \begin{center}
  \hspace*{-0.0cm} \includegraphics[width=12.75cm]{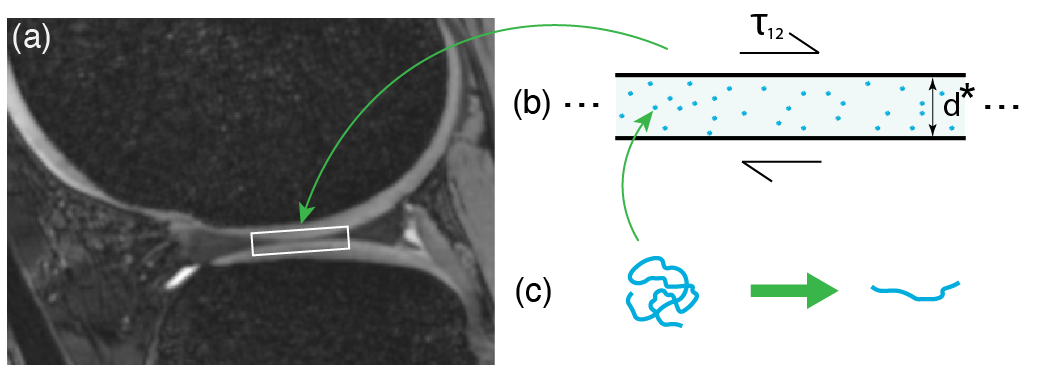}
 \end{center} 
 \caption{The  cartilage-cartilage interface and a schematic of its modelling representation. Plot (a) shows a magnetic resonance image of the knee in the sagittal plane, in which the interface appears as a darker line between the light-coloured cartilage layers within the boxed region of the image. We simplify this interface to a one-dimensional model of shear between flat parallel surfaces in relative motion (b), thus allowing an examination of the large parameter space of plausible synovial fluid properties. In particular, we study the surface tangential shear stress $ \tau_{12}$  for a fluid layer of thickness $d^*$, and the relaxation time and prolateness associated with the molecular weight of the hyaluronan solute, with an increase in prolateness schematically depicted in (c).}
 \label{fig1joint}
\end{figure}

The healthy synovial joint, shown  in Fig.~\ref{fig1joint}a,  has evolved to enable effortless joint motion over decades of extreme physiological loading. Moreover, it provides this load support and lubrication, with minimal wear and   turnover in the opposing cartilage layers. The mechanisms underpinning this function, and the mechanisms by which they break down with disease, are therefore of considerable interest, however the mechanics of the synovial joint has not yet reached a complete understanding.

Nonetheless, it is recognised that the synovial joint has multiple and inter-related lubrication regimes, which include boundary, elastohydrodynamic, boosted, weeping, and mixed-mode mechanisms. These have been  reviewed within a rich literature, for instance: \citet{mccutchen1983,mow1993,dowson2012,ruggiero2020,ateshian2009,jahn2016} and \cite{klein2021}. Differentiating the synovial joint lubrication mechanisms are: the proportion of load bearing of, and contact between, synovial fluid, interstitial cartilage fluid and cartilage solid matrix; the impact of cartilage surface macromolecules; and the extent and direction of fluid transport between the interstitial phase of cartilage and the synovial cavity. 

In boundary lubrication, for example, adhered macromolecules lubricate adjacent cartilage surfaces \citep{jahn2016,klein2021} predicated on the synovial fluid having been forced out from between the cartilage surfaces, as occurs with  extended joint load bearing   \citep{forster1996,forster1999}.  
Further, weeping and boosted lubrication describe the lubrication regimes that emerge with synovial fluid respectively being forced into, or out of, the synovial gap between adjacent cartilage surfaces, in turn altering the prospect of boundary lubrication. It is interesting to note that more recent studies typically support boosted lubrication rather than weeping lubrication \citep{ateshian2009}, further emphasising the importance of  boundary lubrication.

A notable absence from the above list of mechanisms for ameliorating friction in the synovial joint is hydrodynamic lubrication, a popular mechanism in early research \citep{macc1932}. This  would require a thick synovial fluid layer that completely separates  opposing  cartilage surfaces. However, the absence of a local minimum in the Stribeck curve evidences that hydrodynamic lubrication  is not present \citep{gleghorn2008,ateshian2009}. Localised fluid lubrication is, however, present in the mixed-mode lubrication mechanism \citep{seror2015}, where synovial fluid is present between the opposing cartilage surfaces which are also locally sufficiently close to enable localised boundary lubrication  due to surface roughness and asperities.
This also invites the prospect of elastohydrodynamic lubrication, where deformation of the cartilage surface and its asperities due to fluid stress impact the local mechanics \citep{tanner1966}, potentially providing conditions for further localised fluid lubrication \citep{dowson1987}.

Two key questions thus arise in the mechanics of synovial fluid. First, is the timescales under which different modes of lubrication act, and interact. Second, is the influence of synovial fluid material properties on the mechanics of lubrication.
{\bl Addressing the first question requires estimating  the timescale,}
for which boosted lubrication leads to boundary lubrication, prior to which mixed mode lubrication is anticipated to  be present. 
While  \citeauthor{dowson1969}'s (\citeyear{dowson1969}) early estimates indicate a retention timescale of about a minute for the transition from boosted, mixed mode lubrication to boundary lubrication, \citeauthor{wu2017}'s (\citeyear{wu2017}) numerical study of the synovial gap indicate the characteristic retention time in mixed mode lubrication  is in the range of 20-60 minutes due to cartilage surface roughness. 
Hence  mixed-mode lubrication is much more pertinent in synovial joint mechanics than early calculations would indicate and its relevance has also been emphasised  in both experimental and theoretical studies in addition to \citeauthor{wu2017}'s (\citeyear{wu2017}), such as  \citet{katta2008} and \citet{liao2019}. 


{\bl In terms of the second question, the mechanisms} through which the components and dynamics of synovial fluid impact joint motion has been the subject of numerous studies and reviews, including \citet{lai1978,mccutchen1983,tirt1984,mow1992,tamer2013,jahn2016,klein2021}. 
In particular the macromolecule hyaluronan is reported to be the primary determinant of non-Newtonian behaviour for synovial fluid \citep{thurston1978,fam2007}, with shear dependent viscosity and elastic behaviour  \citep{lai1978,tirt1984,fam2007}.   
The variation of synovial fluid viscosity with  shear rate has been subject to significant investigation and has a clear  impact on mechanical behaviour, as investigated by \citet{hron2010}, \citet{bridges2010} and \citet{hasnain2023} for example. 
However, the impact of  synovial fluid elasticity on  mechanical behaviour is not nearly as extensively studied, though expressions for tangential stress are given in the literature for steady shearing in the context of synovial fluid \citep{lai1978}. However, the opposing surfaces in a joint do not move continuously in one direction.

 Here, we simplify the geometry and solid material properties at the cartilage-cartilage interface, {\bl as depicted in Fig.\,\ref{fig1joint}b,} to assess the impact of  elasticity on the tangential stresses, i.e. friction per unit area, that models of synovial fluid are predicted to exert on the cartilage surface for the length and timescales of oscillatory articular joint motion.
Through this investigation we further aim to gain insight into the mechanics of lubrication more generally, and the influence of these mechanics on the degradation process in diseases such as osteoarthritis. 

The wealth of possible models that might be used to consider the non-Newtonian behaviour of synovial fluid is surveyed by \citet{lai1978}. Since elasticity can induce shear dependence in the effective viscosity,  as illustrated by simple calculations \citep{lai1978} for prospective synovial fluid models, we do not consider  shear dependent rheological parameters. 
In turn this isolates effects to the impact of elasticity in this study, as does the neglect of  any compliance in the surfaces confining the synovial fluid in the modelling framework developed below. 
We also do not consider integral constitutive relationships \citep{lai1978} to keep the modelling relatively simple and because we are not aware of synovial fluid modelling using such complex rheology. In considering a differential constitutive relation for synovial fluid rheology, we proceed by utilising the approach of \citet{spagnolie2015}, which develops a Johnson-Segelman model that, in specific parameter limits, presents an Oldroyd-B model \citep{oldroyd1950,hinch2021} noting  that the latter is also a model of focus for synovial fluid  in \citet{lai1978}.

In Section \ref{sec2} we will present the model of elasticity that we will use, highlighting its similarities and generalisations from the Oldroyd-B model, as well as its physical motivations and limitations. Concomitantly, we will derive the 
governing equations for a simple shear test  on the length and timescales associated with a joint to investigate how elasticity impacts the tangential stress, and thus friction, exerted by the fluid on its confining surfaces. 
Given the uncertainty and range of physical parameter values, as further detailed in modelling development, the simplest setting for investigating the modelling framework is sought. Hence, in Section \ref{sec3} we consider the impact of surface fluctuations to justify the use of model simplifications associated with flat confining surfaces during oscillatory joint motion. 
The resulting model simplification is exploited in Section \ref{sec4} to enable a mathematical investigation, which gives rational and accurate approximations of the tangential stress for oscillating shear {\mg associated with joint motion.}
These expressions are tested via numerical computation, before drawing conclusions  across the relevant parameter space in the concluding section {\mg concerning  the impact of   elasticity on tangential stress and thus friction} for flow on the scales of oscillatory joint motion.

\section{The elasticity model and the  governing equations}
\label{sec2} 

 Below we develop a constitutive relation, and the mechanics, for synovial fluid by considering a viscous and elastic polymer solute, in particular hyaluronan, within a solvent that is essentially  water, albeit with physiological ions. The resulting governing equations are subsequently derived for flow  at finite strain, within a confining geometry that possesses a depth that is representative of the synovial gap between opposing cartilage surfaces.  


\subsection{The constitutive relation and elasticity}\label{sec201}

Following the strategy of Section 4 within \citet{spagnolie2015}, we take the polymer solute to behave as the simplest incompressible  fluid with both viscous and elastic behaviour, that is, a Maxwell fluid. Thus combining the mechanics of this solute with a  Newtonian fluid solvent, the Cauchy stress $\bm\sigma^{*}$ is given by 
\begin{eqnarray} \label{cst} \sigma^*_{ij} = - p^*\delta_{ij}+\tau^*_{ij}+\tau^{s*}_{ij}.
\end{eqnarray} 
Throughout, asterisks denote dimensional quantities and subscripts refer to indices for inertial reference frame Cartesian components.  The pressure 
 $p^*$ acts as a Lagrange multiplier that enforces incompressibility, while   $\bm\tau^*, ~\bm\tau^{s*}$ are respectively the solute and solvent contributions to the deviatoric stress. Proceeding,  let $u^*_i$ denote the $i^{\tiny{\mbox{th}}}$ inertial frame Cartesian component of the velocity field,
and let the subscript $,i$ denote a partial derivative with respect to $x^*_i$, the $i^{\tiny{\mbox{th}}}$ inertial frame Cartesian coordinate. Then, with  the use of summation convention and $\mu^*,\lambda^*$ giving the polymer solute viscosity and elastic  relaxation time,  the constitutive relations between the velocity field and the deviatoric stresses are given by \citep{spagnolie2015,hinch2021} 
\begin{eqnarray} \nonumber 
& \dot{{\cal E}}^*_{ij} = \dfrac 1 2 (u^*_{i,j}+u^*_{j,i}) , ~~~~~~
\Omega^*_{ij} =  \dfrac 1 2 (u^*_{i,j}-u^*_{j,i}) , ~~~~~~  \tau^{s*}_{ij} = 2\mu^{s*} \dot{{\cal E}}^*_{ij} = \tau^{s*}_{ji}, & \\ \label{ge1} 
&  \tau^*_{ij} +\lambda^* \dfrac{D^{JS}\tau^*_{ij}}{Dt^*}= 2\mu ^*\dot{{\cal E}}^*_{ij},  & \\ & 
\dfrac{D^{JS}\tau^*_{ij}}{Dt^*} := \dfrac{\partial \tau^*_{ij}}{\partial t^*} + u^*_p \tau^*_{ij,p} -\Omega^*_{ip}\tau^*_{pj}+\tau^*_{ip}\Omega^*_{pj}-a\left(\dot{{\cal E}}^*_{ip} \tau^*_{jp} + \tau^*_{ip}  \dot{{\cal E}}^*_{pj}  \right).&  \nonumber 
\end{eqnarray} 
Here, the superscript $JS$ denotes the frame-invariant Johnson-Segelman derivative \citep{spagnolie2015,hinch2021}, $t^{*}$ represents time, $\Omega^*_{ij}$ is the local vorticity, $ \dot{{\cal E}}^*_{ij} $ is the local rate of strain, and  $\tau^*_{ij}=\tau^*_{ji}$ is assured by angular momentum conservation.  Note that $\lambda^*$ measures the extent of elasticity for the polymer solute, and that the solvent contribution to the deviatoric stress is that of a Newtonian viscous fluid of viscosity $\mu^{s*}$. 
 
   The Johnson-Segelman derivative tracks the movement and reorientation of the polymer solute and is the most general frame-invariant material derivative for a rank two tensor. Thus the free dimensional parameter $a$,  referred to as the slip parameter by \cite{spagnolie2015}, cannot be restricted without further appeal to physics or experiment. In this regard, \citet{hinch2021} helpfully recall  \citeauthor{jeffrey1922}'s (\citeyear{jeffrey1922})   result that spheroids in linear flow rotate with all of the local vorticity flow and a proportion $(r^2-1)/(r^2+1)$ of the local rate of strain, 
where $r\in[0,\infty]$ is the spheroid aspect ratio. Hence,  $r\in\{0,\infty\}$ corresponds to limiting cases, respectively of  very oblate spheroids  and very prolate spheroids, while $r=1$ corresponds to a solute of spherical particles. Noting that the additional contributions to the invariant derivatives are  tracking the reorientation of the solute,  \citet{hinch2021} then proceed to suggest that one should take 
$$ a = \dfrac{r^2-1}{r^2+1} \in[-1,1],$$
while keeping a unit coefficient for the vorticity contribution in selecting the invariant derivative for solutes consisting of  spheroids. 
Hence we limit the slip parameter  to the interval $[-1,1]$. In particular,    $a=-1$ corresponds to  a lower convected derivative, $a=0$ the co-rotational  Jaumann derivative and   $a=1$ corresponds to an upper convected derivative, {\bl with increasing $a$ corresponding to increasing prolateness of the solute, in particular hyaluronan, as schematically indicated in Fig.\,\ref{fig1joint}c.}

The Oldroyd-B model suggested by \citet{lai1978} is generated by the above framework for the limiting case $a=1$, once the solvent viscosity is taken into account by rewriting the constitutive relation in terms of $ \bm \tau^* +\bm\tau^{s*}$, the total deviatoric stress. The Oldroyd-A model is analogously generated with the lower convected derivative, $a=-1$, but is far less common as its behaviour is typically not observed \citep{hinch2021}. 
The Maxwell model, corresponding to the limiting case of $\mu^{s*}=0$,  is also one of the many cases  considered by \citet{lai1978}, who found a point-estimate data fitting yielded a best fit of $a= 0.97$. In turn, this might suggest that an upper convected derivative, and thus the Oldroyd-B model, as a suitable model. However,  as we shall see below, the tangential stress dependence on the slip parameter $a$ is sensitive close to $a=1$.  
Furthermore, hyaluronan molecular weight decreases with disease \citep{band2015} 
while  hyaluronan is predicted to be capable of coiling at higher molecular weights \citep{taweechat2020} in turn influencing its prolateness and  thus possibly the slip length $a$ given the latter's physical interpretation. Hence,  we do not restrict the study to $a=1$.

\subsection{The dimensional equations for flow on joint length and timescales}\label{sec21} 

To consider dynamics  on joint length and timescales, we  consider the model synovial fluid moving through a narrow Cartesian channel. Let $x^*$ denote the Cartesian coordinate along the channel, which is of the scale $d^*$, with 
\begin{eqnarray}\label{geom} 
{\bl y_1^*(x^*) \leq y^* \leq  d^*+y^*_2(x^*,t^*),}~~~~~~~~ |y_1^*|, ~|y_2^*| \ll d^*. 
\end{eqnarray} 
The boundary at $y^* = y_1^*(x^*)$ is fixed, and the 
boundary at {\bl $y^* = d^*+ y_2^*(x^*,t^*)$ moves  relative to the boundary at 
$y^* = y_1^*(x^*)$.}  
The remaining bulk equations are those for the conservation of momentum, angular momentum and mass. Combining  the latter with incompressibility, we thus have 
\begin{equation} \rho^* \left( \frac{\partial u^*_{i}}{\partial t^*} + u^*_p u^*_{i,p}\right) = \sigma^*_{ij,j},~~~~ \tau^*_{12}=\tau^*_{21},~~~~u^*_{i,i}=0=\dot{\cal E}^*_{ii},  \label{ge2}
\end{equation} where $\rho^*$ is the fluid density, assumed constant.  
For initial conditions we assume the fluid has started from rest and thus 
\begin{equation} \label{Ics0} 
(u^*(x^{*},y^*,0),~v^*(x^{*},y^*,0))=\bm 0 , \qquad \bm\tau^*(x^{*},y^*,0)=\bm 0,
\end{equation} with  $p^*$ constant at the initial time, though it is not zero as its hydrostatic contribution  supports the load on the joint at rest (outside of boundary lubrication, which is not in scope).  

The model thus far consists of the constitutive relation of Eqs.~\eqref{cst},\eqref{ge1}, 
the domain Eq.~\eqref{geom}, the conservation equations \eqref{ge2},
and the initial conditions, Eq.~\eqref{Ics0}.
The boundary conditions will be specified after further discussion in later sections. The fluid is assumed to be driven by the relative motion of the joint surfaces;  we consider the prospect of  treating the joint surfaces as smooth and flat, with $y_1^*(x^*)=y_2^*(x^*,t^*)=0$ in Eq.\,\eqref{geom} and thus a constant separation between the two cartilage layers. This enables very extensive model simplification in determining a one-dimensional model below. However,  the prospect of utilising a smooth and flat surface for model simplification first requires motivation and thus in the next section we briefly consider the impact of a surface fluctuation in the channel geometry.

\section{The impact of surface fluctuations. The microscale model.}\label{micsc}
\label{sec3}

In this section the objective is to illustrate that the model predictions for velocity profiles,  and hence the associated stresses,  are robust to  surface fluctuations at the synovial-fluid cartilage interface.  As the focus is on the impact of small scale  fluctuations 
we use a periodic surface to enable an expedient numerical solution of  the full set of equations on a relatively small spatial domain. Hence the model is at a microscale and designed to illustrate the impact of surface fluctuations rather than to completely mimic a joint.  
In particular, we consider a half-depth of the synovial fluid layer with 
$$ y_1^*(x^*) \leq y^* \leq  d^*/2,~~~~~~~~ |y_1^*| \ll d^*, ~~~~ \bm u^*(x^*,y^*_1(x^*),t^*)=\bm 0,$$ 
and a boundary condition of
$$ \bm u^*\left(x^*,\frac{d^*}2,t^*\right)= U^*(t)\bm e_x.$$ Thus, there is a {\bl pure} shearing motion imposed at the channel midplane by the boundary condition, with no slip at the lower channel boundary, $y^*=y^*_1(x^*)$.

\begin{table} 
  \begin{center}
    \begin{tabular}{lll}
     \noalign{\smallskip} 
      Param & Estimate or value & Reference or justification
      \\
      \noalign{\smallskip} \hline \noalign{\smallskip}
      $a$ & $ [-1,1]$  & Slip parameter in the invariant derivative, see Section \ref{sec201}. \\
      $a$ & $ 0.972$  & Point estimate in \citet{lai1978}.
      \\ $\lambda^{*}$ &  [0.65~s, 5.8~s]& Relaxation time for synovial fluid \citep{fam2007}.  \\ $\lambda^{*}$ &    4.5~s & Relaxation time for synovial fluid \citep{Rwei2008}.
      \\ $\lambda^{*}$ & 0.02~s, 0.42~s & Relaxation time for synovial fluid \citep{lai1978}.
      \\ $\lambda^{*}$ &  0.02~s, $<0.02$~s& Relaxation time for RA synovial fluid (seronegative,   \\   & &    seropositive). \citet{fam2007}. 
      \\  $T^{*}$ & 4.5~s & Non-dimensionalisation time. Typical relaxation time  \\
      &  & for hyaluronic acid. \\
            $d^{*}$ &  $ 10^{-7}$m$-1.5\times10^{-6}$m  & Separation of adjacent cartilage layers, from calculations \\
      & & by \cite{tanner1966} and others; see \cite{torzilli2021}. \\
      $W^{*}$ & $0.01$~m~s$^{-1}-0.1$~m~s$^{-1}$ & Relative speed of adjacent cartilage layers. \\
      $\mu^{s*}$ & $7\times 10^{-4}$~N~m$^{-2}$s ~& Shear viscosity,  water at body temperature. \\
      $\mu^{Syn^*}$ & $3-30$~N~m$^{-2}$s & Shear viscosity, synovial fluid  at $\approx$1~Hz. \\
      & & (\cite{bingol2010},  \cite{schurz1987}). \\
      $\mu^{Syn^*}$ & $0.1$~N~m$^{-2}$s & Shear viscosity, synovial fluid (osteoarthritis)  at $\approx$1~Hz \\ 
       &   &  \citep{schurz1987}. \\
      $\mu^{*}$ & $\mu^{Syn^*}$ & Shear viscosity,  solute contribution at $\approx$1~Hz. See caption. \\   
      $\rho^{*}$ & $10^3$ kg m$^{-3}$  & Density of Synovial fluid $\approx$ density of water. \\
      & &   (Rabbit measurements, \citet{knox1988}).\\ 
    $\nu^*$ & $0.5-2$~Hz &  Common frequency of periodic joint movement.\\
   $ T^{u^*} $& $  T^{u^*}=1/\nu^* $ &   Common timescale  of periodic joint movement.\\
      \noalign{\smallskip}  
    \end{tabular}
  \end{center}
  \caption{The values of dimensional parameters and the non-dimensional slip parameter, $a$.
  The estimates of $a=0.972, \lambda^*=0.02$~s are from a point-estimate fitting a Maxwell model to bullock synovial fluid, but fitting an Oldroyd-B model to human synovial fluid gave $\lambda^*=0.42$~s \citep{lai1978}.  This latter  fitting is incomplete and does not estimate $a$ for example.  The smaller value of $\lambda^*\approx 0.02$~s is more indicative of measurements of  synovial fluid in Rheumatoid Arthritis (RA) \citep{fam2007}.  
  Synovial fluid is shear-thinning and thus its viscosity reduces on increasing the frequency of the forcing and our human synovial viscosity parameter estimate is taken for this forcing on the scale of ord(1)~Hz.  Given $\mu^{Syn^*}\gg \mu^{s^*}$ we have the viscosity of the synovial fluid is much greater than that of its solvent and hence we approximate the solute contribution to the total viscosity, $\mu^{Syn^*}$, to be $\mu^*\approx \mu^{Syn^*}$.     
  }
  \label{Tab:Pars}
\end{table}

For definiteness, we take the spatial period of the surface fluctuation to be $d^*$, with an amplitude of $\eta d^*$, with $0<\eta\ll 1.$
To proceed with the microscale model and the demonstration that $y_1^*(x^*)$  does not have a substantial impact,  we non-dimensionalise Eqs.~(\ref{ge1}) using the scalings
\begin{alignat*}{7}
\mathbf{x}^{*}&=d^{*}\mathbf{x}, ~~~~~~ & t^{*}&=T^{*}t, ~~~~~~ & \lambda^{*}&=T^{*}\lambda, ~~~~~ & \mathbf{u}^{*}&=W^{*}\mathbf{u}, ~~~~~~~
&\bm\tau^{*}&=\frac{\mu^{*}W^{*}}{d^{*}} \bm\tau, \\ \bm\dot{\cal{E}}^{*}&=\frac{W^{*}}{d^{*}} \bm\dot{\cal{E}}, ~~~~~~ & \bm\Omega^{*}&=\frac{W^{*}}{d^{*}} \bm\Omega, ~~~~~~ & p^{*} &= \frac{\mu^{*}W^{*}}{d^{*}} p, ~~~~~
&\sigma^{*} &= \frac{\mu^{*} W^{*}}{d^{*}} \sigma,
\end{alignat*}
where $T^*,W^*$ are representative of the timescale and magnitude of the velocity, and obtain the dimensionless equations
\begin{eqnarray}\nonumber 
    \dot{\cal{E}}_{ij} &=& \frac{1}{2} \left( u_{i,j} + u_{j,i} \right), ~~~~~~~~~~~\,
    \Omega_{ij} = \frac{1}{2} \left( u_{i,j} - u_{j,i} \right), ~~~~~~~~~~~~~~~
    \tau^{s}_{ij} = 2 \frac{\mu^{*}_{s}}{\mu_{\text{syn}}^{*}} \dot{\cal{E}}_{ij}, \\ \nonumber 
 2\dot{\cal{E}}_{ij}  &=&   \delta \tau_{ij} + \lambda \left( \delta \frac{\partial \tau_{ij}}{\partial t} + u_{p} \frac{\partial \tau_{ij}}{\partial x_{p}} - \Omega_{ip}\tau_{pj} + \tau_{ip}\Omega_{pj} - a \left( \dot{\cal{E}}_{ip} \tau_{pj} + \tau_{ip} \dot{\cal{E}}_{pj} \right) \right), ~~~~~~~\\ 
    \sigma_{ij} &=& -p \delta_{ij} + \tau_{ij} + \tau^{s}_{ij}, ~~~~~~~
 \frac{\partial \sigma_{ij}}{\partial x_{j}} =   \text{Re} \left( \delta \frac{\partial u_{i}}{\partial t} + u_{p} u_{i,p} \right), ~~~~~~~~
    u_{i,i} = 0,  \label{microeqns} 
\end{eqnarray}
with $\delta= d^{*}/(W^{*}T^{*})$ and other parameters discussed in Table~\ref{Tab:Pars}.

\begin{figure}
\begin{center}
    \includegraphics[width=10cm]{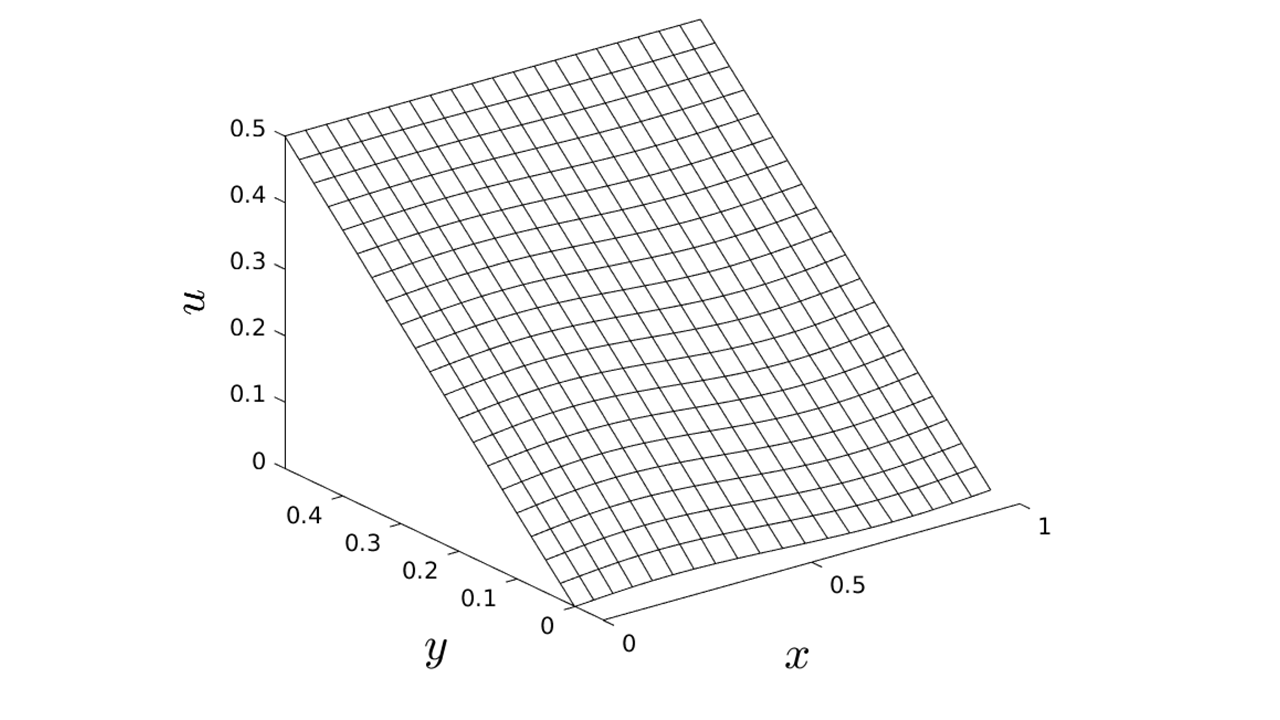}
\end{center}
\caption{The velocity $u$ obtained from the microscale model at a snapshot in time when the driving velocity at $y=1/2$ in the non-dimensional microscale model is at its largest and any initial transients have decayed. The non-dimensional parameters for the simulation are given by $\delta=d^*/[W^*T^*]=10^{-5},$ a typical value given the constraints on the scales, a non-dimensional elastic relaxation time of $\lambda=1.0,$ and a surface fluctuation amplitude of $\eta=0.01$. In addition, the 
the slip parameter of the constitutive relation {\bl is given by $a=0$, as an exemplar value of the slip parameter that gives neither an upper convected nor a lower convected Johnson-Segelman derivative.} Note that the profile is essentially that of shear flow with translation invariance in the $x$-coordinate, which is the profile seen when $\eta=0$, {\bl as may deduced from the results presented below. Hence,} the surface fluctuation has minimal impact on the solution, illustrating that using a constant surface profile for the one-dimensional model of the next section is a reasonable approximation.}
\label{Fig:Micro}
\end{figure}

To assess the effect of perturbations in the boundary to the solution of the dimensionless equations we take 
$$y^*_1(x^*) = \eta d^* \sin(2\pi x^*/d^*),$$ so that we  solve these equations for the domain 
\begin{equation} \label{microbc1} 
0<x<1, ~~~~~~ \eta \sin(2\pi x)<y<0.5, 
\end{equation} 
with periodic boundary conditions in the $x$-direction and 
\begin{equation} \label{microbc2}
\bm u = \bm 0   ~~~~\mbox{on}~~~~ y=\eta\sin (2\pi x), ~~~~~~ \bm u=0.5 \sin(2\pi t ) \bm e_x, ~~~\mbox{on}~~~y=0.5~.
\end{equation}
Note that as the normal velocity is zero on both upper and lower boundaries the terms  $  u_p \tau_{ij,p} $  
do not generate the need for stress boundary conditions on the upper and lower domain boundaries. 

Thus the model is given by the equations Eqs.\,\eqref{microeqns}, with the trivial initial conditions of Eq.\,\eqref{Ics0}, lateral periodic boundary conditions and the boundary conditions Eqs.\,\eqref{microbc1},\eqref{microbc2}. A solution  is presented  for 
$u$  in Figure~\ref{Fig:Micro} at a time given by 
$t=10.25$, the value of $t$ where $u$ takes its maximum value on $y=0.5$. The parameters used prior to non-dimensionalisation are given in the figure caption and representative of the possibilities in Table~\ref{Tab:Pars}.

{\bl One can immediately observe that shear flow is an excellent approximation to the solution. Given a shear flow might be expected in the case of flat surfaces, 
with  $y^*_1(x^*)=y^*_2(x^*,t^*)=0$, we impose this as an ansatz below. 
This then  allows an increase in  the lateral lengthscale of the modelling and the  development of a  one-dimensional  model amenable to mathematical investigation. In the early stages below it is mathematically confirmed that a shear flow emerges, and this is uniquely given with the specification of the relative movement of the surfaces, confirming the self-consistency of the ansatz. This emphasises that one may self-consistently neglect small surface fluctuations to good approximation, as a shear flow driven by the relative surface movement essentially emerges when surface fluctuations are small, and the shear flow is that of surfaces in relative motion in the absence of the fluctuations. In turn, the  simplifications 
from  neglecting small surface fluctuations are  especially  useful as they enable a mathematical summary of parameter dependencies in the generation of tangential stress, and thus friction, across the large space of plausible parameters for synovial fluids.}

\section{Modelling Joint Motion. One Dimensional Model.}\label{sec4} 

In the one dimensional model, the prospect of eliminating the   lateral variation is considered 
and hence the non-dimensionalisation will be reconsidered below. In the previous section we  motivated the  neglect of any surface geometry fluctuation at the synovial-cartilage interface, i.e.\ $y_1^*(x^*)=y_2^*(x^*)=0$ in Eq.\,\eqref{geom}, so that the geometry is  given by 
\begin{equation}\label{geom1} 
 0< y^* <d^* 
\end{equation}
for all $x^*$. Hence the upper and lower velocity boundary conditions, assuming the lower cartilage is stationary, will reduce to 
\begin{equation}\label{macrobc1} 
 \bm u^*(x^*,0,t^*) = \bm 0, ~~~~    \bm u^*(x^*,d^*,t^*) = U^*(t^*) \bm e_x  .
\end{equation}

We also assume that the surface does not compress and extend so that there is no 
$x^*$-dependence in the velocity boundary conditions. In turn this means  that no aspect of the model has variation in this direction. Hence we no longer have lateral boundary conditions  and, in addition,  by incompressibility and the modelling assumption that there is negligible fluid motion through the lower boundary of the channel at $y^*=0$,  we have 
$$ v^*(x^*,y^*,t^*)= \int_0^{y^*} \dfrac{\partial u^*}{\partial x^*}(x^*,\bar{y}^*,t)~\mathrm{d}\bar{y}^* = 0.$$ Hence the flow field is of the much simpler form 
\begin{equation}\label{ufield} 
\bm u^*=(u^*(y^*,t^*),0),
\end{equation}
which automatically satisfies the incompressibility requirement and removes variation in the lateral direction. In addition, this entails that $\bm \dot{\cal E}^*, ~\bm \Omega^*$ are proportional to $\partial u^*/\partial y^* $ and thus are scaled using the inverse lengthscale $1/d^*$ below, as the representative scale of spatial derivatives.  The absence of vertical fluid velocity and the $x^*$-independence in the model ensures that the 
 terms $ u^*_p \tau^*_{ij,p} $ are identically zero and thus there are no spatial derivatives of the stress; hence    boundary conditions for $\tau^*$ are not required.

Typical dimensional values for the parameters considered at the joint scale are presented in Table \ref{Tab:Pars}. 
With  $W^*$   representative of the velocity in the $x$-direction, and 
$ T^{*}=4.5$~s  representative of the typical elastic relaxation timescale, as given in Table \ref{Tab:Pars},  
another lengthscale in the model is $W^*T^*$. 
Anticipating  the stress need not diverge as layers become ever  thinner we consider the non-dimensionalisation using this lengthscale for the stress: 
\begin{equation}\label{nondim}  \bm \tau^{*} = \frac{ \mu^{*}W^*}{T^{*}W^*} \bm \tau= \frac{ \mu^{*}}{T^{*}} \bm \tau. 
\end{equation}
We then use the remaining rescalings 
\begin{alignat}{9} \nonumber
  t^{*} & = T^{*} t, \qquad & 
  y^{*} &= d^{*} y,  \qquad & 
  \bm \dot{\cal E}^{*} &= \frac{W^{*}}{d^{*}} \bm\dot{\cal E}  ,  & \qquad
  \bm\Omega^* &= \frac{ W^{*}}{d^{*}} \bm\Omega, & \\
  p^{*} & = \dfrac{\mu^*}{T^*} p  
  \qquad &  u^{*} & = W^{*} u, \qquad & 
  \lambda^{*} &= T^{*} \lambda,  & 
  U^{*} &= W^{*} U,  &  \label{lam}
\end{alignat}
to non-dimensionalise, where the pressure-scaling  is taken from the scaling of the stress, and the resulting non-dimensional parameters are given in Table $\ref{tab2}$. However, we do not set the non-dimensional viscoelastic relaxation time $\lambda$ to be identically unity so that the impact of changes in elastic relaxation timescales can be easily tracked.  



Note  $u_pu_{i,p}$ is identically zero from Eq.~\eqref{ufield}, and the dimensionless rate-of-strain and rotation tensors may be  written as 
\begin{eqnarray*}
  \bm \dot{\cal E} = \frac{1}{2} \frac{\partial u}{\partial y} \begin{pmatrix} 0 & 1 \\ 1 & 0 \end{pmatrix}, \qquad \bm \Omega = \frac{1}{2} \frac{\partial u}{\partial y} \begin{pmatrix} 0 & 1 \\ -1 & 0 \end{pmatrix}.
\end{eqnarray*} 
Recalling $x$-derivatives are null, the 
non-dimensionlisation of the conservation of momentum  equation thus gives 
\begin{eqnarray}\label{gend00} 
  \mbox{Re} \dfrac{\partial u}{\partial t} = \frac{\partial \tau_{12}}{\partial y}+ \zeta \frac{\partial^2 u}{\partial y^2} ,~~~~ ~~ 0=-\frac{\partial p}{\partial y}+\frac{\partial \tau_{22}}{\partial y},
\end{eqnarray} 
with 
\begin{eqnarray}\label{prm12} \mbox{Re}  &=& \dfrac{\rho^*W^*d^*}{\mu^*} \in [5\times 10^{-9},\, 
1.5\times10^{-6}] , \\ \zeta &=& 
\dfrac{\mu^{s*}}{\mu^*}\frac{W^*T^*}{d^*}
\in[0.7,\, 2\times10^{5}].~~~~~~~~
\end{eqnarray} 
Thus, while solvent viscosity is often neglected this is not legitimate a priori here, with $\zeta$ elevated due to the aspect ratio of the thin synovial film.

\begin{table} 
  \begin{center}
    \begin{tabular}{llll}
     \noalign{\smallskip} 
      Param & Definition & Range & Comments
      \\
      \noalign{\smallskip} \hline \noalign{\smallskip}
      Re &  $\dfrac{\rho^*W^*d^*}{\mu^*}$   & $[5\times 10^{-9},\, 
1.5\times10^{-6}] $& The Reynolds number. Eq.~\eqref{prm12}. \\ ~~\\ 
Re$_u$ &  $Re\dfrac{T^*}{T^{u*}}$   &  $[ 10^{-8},\, 
1.4\times10^{-5}] $ & Velocity timescale  Reynolds number. Eq.~\eqref{ret}. \\ ~~\\
      $\zeta$ & $\dfrac{\mu^{s*} W^*T^*}{\mu^* d^*} $ & $ [0.7,\, 2\times10^{5}]$   &  Impact of solvent viscosity. Eq.~\eqref{prm12}. \\~~\\
      $\delta$ & $\dfrac{d^*}{W^*T^*} $ & $ [2\times10^{-7},3\times10^{-5}]$   &  Fluid depth / lateral 
   lengthscale. Eq.~\eqref{prm3}.\\~~\\
      $\nu$ & $\nu^*T^* $ & $ [2.25,9]$   &  Non-dimensional joint motion frequency. Table \ref{Tab:Pars}.\\~~\\
      $\lambda$ & $\lambda^*/T^* $ &  $[0.1,1.3]$  &   Non-dim viscoelastic relaxation time\, Eq.~\eqref{lam}. \\~~\\
      $\lambda$ & $\lambda^*/T^* $ &  $\leq 4.4\times 10^{-3}$  &   Non-dim viscoelastic relaxation time for RA.  \\~~\\
      $\alpha $ & $\lambda(1-a^2) $ &    &   Auxiliary parameter in ODEs, Eqs.\,\eqref{odes}. \\ ~~\\
      $\beta$ & $\lambda^2(1-a^2) $ &    &   Auxiliary parameter in ODEs, Eqs.\,\eqref{beta}. \\
      \noalign{\smallskip}  
    \end{tabular}
  \end{center}
  \caption{Non-dimensional parameter values. A number of parameter groupings arise in deducing the non-dimensional model, and are summarised above. A range of $\lambda$ is given for commonly reported elastic relaxation times for synovial fluid in the absence of disease, with the lower value of $4.4\times 10^{-3}$ or less,  reported for rheumatoid arthritis, RA (treating the lower estimate of \citet{lai1978}  as  within the diseased range).  The auxiliary parameters, $\alpha,~\beta$, used in the modelling are also summarised, noting the parameter $a$ is summarised in Table \ref{Tab:Pars}.}
  \label{tab2}
\end{table}

The timescale on which the velocity changes, which we denote by $T^{u*}$, does not have to be on the same timescale of the elastic relaxation, $T^*\sim \lambda^*$, but instead corresponds to the timescale on which $U^*(t^*)$ changes, that is the relative motion of the boundaries due to joint movement, which can be at a frequency of $$\nu^* =\frac 1{T^{u*}} \approx 0.5-2 ~\mbox{Hz}$$  for walking to jogging for example. Thus the scale of the velocity term above when the flow is driven by the relative motion of the boundaries is more accurately represented by 
\begin{equation}\label{ret}  \mbox{Re}_u=\mbox{Re} \frac{T^*}{T^{u^*}} \in [7\times 10^{-8},\, 8\times10^{-5}].
\end{equation}
Thus, it would take unphysiological joint motion frequencies of  more than one  kilohertz for this to approach a magnitude of unity, and become relevant.  An exception would be if the non-dimensional initial conditions for  $\tau_{12,2}(0,y)$ and $u_{yy}(0,y)$ were such that 
\begin{equation}\label{icin} \frac{\partial \tau_{12}}{\partial y}(0,y)+ \zeta \frac{\partial^2 u}{\partial y^2}(0,y) \not = 0 ,
\end{equation}
 in which case the Re$\, \partial u/\partial t$ would have to be non-zero initially, ensuring initial evolution on a very rapid timescale $t/\mbox{Re}$, with the time-scaling to ensure the inertial term did not remain trivial.
Given the initial conditions we use are those at rest, with zero deviatoric  stress, Eq.~\eqref{icin}   does not apply and hence the prospect of inertia-driven temporal boundary layers are not considered.

As a result, henceforth we set Re$\,=\,$Re$_u=0$, corresponding to the  neglect of inertia, and the conservation of momentum equations become 
\begin{eqnarray}\label{gend0} 
0=\frac{\partial \tau_{12}}{\partial y}+\zeta \frac{\partial^2 u}{\partial y^2} ,~~~~ ~~ 0=-\frac{\partial p}{\partial y}+\frac{\partial \tau_{22}}{\partial y}.
\end{eqnarray}


The dimensionless constitutive relation reduces to 
\begin{eqnarray}\label{gend2}
 \delta   \bm  \tau + \lambda \left[  \delta 
  \frac{\partial \bm \tau}{\partial t} -\ \bm \Omega \bm \tau +  \bm \tau \bm \Omega -  a ( \bm \dot{\cal E} \bm \tau + \bm\tau \bm \dot{\cal E} )   \right] 
= 2 \bm \dot{\cal E}, 
\end{eqnarray}
where  
\begin{eqnarray}\label{prm3}
 \delta = \dfrac{d^*}{W^*T^*} \in[2\times10^{-7},3\times10^{-5}].
\end{eqnarray}
However, while $\delta \ll 1$, it cannot be simply set to zero. For example, if the fluid was Newtonian, the non-dimensional solution between moving plates  would simply be  shear flow with  
$$ u^* = \frac{yW^*}{d^*} U(t), $$
 where $U(t)$ is the non-dimensional relative velocity of the plates. Hence, the flow field in the limit $d^*\rightarrow 0$,  $\delta \rightarrow 0$ is singular, due to an incompressible flow with a finite flux within a region that has its volume  tending to zero. This aspect of the physics is not related to the presence or absence of fluid elasticity and thus we retain $\delta >0$ for the consideration of viscoelastic fluids below.

Proceeding, with the retention of $\delta$ and recalling $\tau_{12}=\tau_{21}$ by angular momentum conservation, we thus have 
\begin{eqnarray}\label{gend30}
  \tau_{11} + \lambda 
  \frac{\partial \tau_{11}}{\partial t} -   \frac \lambda  \delta (1+a)    \frac{\partial u}{\partial y}\tau_{12} 
 &=&  0 \\ \label{gend31}  \tau_{22} + \lambda 
  \frac{\partial \tau_{22}}{\partial t} +   \frac \lambda  \delta (1-a)    \frac{\partial u}{\partial y}\tau_{12} 
 &=&  0 \\ \label{gend32} \tau_{12} + \lambda 
  \frac{\partial \tau_{12}}{\partial t} +   \frac \lambda  {2\delta}   \frac{\partial u}{\partial y}\left[ (1-a)  
 \tau_{11}  -  (1+a)    \tau_{22} \right]  &=& \frac 1 \delta \frac{\partial u}{\partial y}. 
\end{eqnarray}
Note that in the limit that the viscoelastic relaxation time reaches zero, $\lambda\rightarrow 0$,  we have $\tau_{11}=\tau_{22}=0$ and 
$$ \tau_{12} =  \frac 1 \delta \frac{\partial u}{\partial y},$$ 
and thus a non-dimensional Newtonian behaviour of the stresses. Finally, the boundary and initial conditions are trivially inherited,
\begin{eqnarray}\label{bcic} 
& u (t ,y =0)  =0, \quad  u (t ,y =1)=U(t), \quad 
u (t=0,y )= 0 , \quad \bm \tau (t=0,y )=\bm 0, ~~~~~~~~~\, &
\end{eqnarray}
with $U(0)=0$ for consistency between the initial and boundary conditions. We are  particularly interested in the tangential stress, $\tau_{12}$, at the upper and lower surfaces, $y=0,1$, which dictates the  predicted level of friction at the synovial-cartilage interface. 

In summary,  the first of Eqs.~\eqref{gend0}, and Eqs.~\eqref{gend30}-\eqref{gend32} gives four scalar equations for the four scalar unknowns, $u,~ \tau_{12}, ~ \tau_{11}, ~ \tau_{22}, $ while 
the second of Eqs.~\eqref{gend0} gives the pressure up to an additive function of time, which we comment further on below.  The model is then closed by the boundary and initial conditions of Eq.~\eqref{bcic}.

\subsection{Model solution}  

 We first of all define 
 $\chi^{-},\chi^{+}$ via 
\begin{eqnarray}\label{chidef}
    \chi^{-} = \frac{1}{2} \left( (1-a) \tau_{11} - (1+a) \tau_{22} \right), \quad \chi^{+} = \frac{1}{2} \left( (1-a) \tau_{11} + (1+a) \tau_{22} \right). 
\end{eqnarray}
From Eqs.~\eqref{gend30}-\eqref{gend31} we see that $\chi^{+}$ decouples, allowing us to eliminate one equation.
Further $\chi^{+}$ satisfies 
\begin{eqnarray}
    \lambda \frac{\partial \chi^{+}}{\partial t} + \chi^{+} = 0,~~~~ \chi^+ =\chi^+(t=0)\mathrm{e}^{-t/\lambda}, \label{chiplus}
\end{eqnarray}
and given the initial conditions are zero stress, this collapses to  $\chi^{+}=0$.
The remaining equations are 
\begin{eqnarray}
    \lambda \frac{\partial \chi^{-}}{\partial t} + \chi^{-} - \frac \lambda \delta (1-a^{2}) \frac{\partial u}{\partial y}  \tau_{12} &=& 0, \label{gend40} \\
    \lambda \frac{\partial \tau_{12}}{\partial t} + \tau_{12} + \frac \lambda \delta \frac{\partial u}{\partial y} \chi^{-} &=&  \frac 1 \delta \frac{\partial u}{\partial y} .\label{gend41}
\end{eqnarray}
Integrating the first equation in Eq.~(\ref{gend0}) we also have 
\begin{eqnarray}
    \tau_{12} + \zeta \frac{\partial u}{\partial y} = f(t), \label{Eq:ODE1}
\end{eqnarray}
  where $f(t)$ is a  function of time that ultimately can be determined.  Using Eq.~(\ref{Eq:ODE1}) to eliminate $\partial u/\partial y$ from Eqs.~(\ref{gend40}) and (\ref{gend41}) gives
  \begin{eqnarray*}
      \zeta \left( \chi^{-} + \lambda \frac{\partial \chi^{-}}{\partial t} \right) - \frac{\lambda \tau_{12}}{\delta} \left( 1-a^{2} \right) \left(f-\tau_{12} \right) &=& 0, \\
      \zeta \left( \tau_{12} + \lambda \frac{\partial \tau_{12}}{\partial t} \right) + \frac{1}{\delta} \left( \lambda \chi^{-}-1 \right) \left(f - \tau_{12} \right) &=& 0.
  \end{eqnarray*}
  As there is no $y$-dependence in the initial conditions for $\chi^{-},\tau_{12}$ and all other quantities are constant, there is no $y$-dependence on the solution as $t$ evolves, and so the system of equations above is a system of ODEs in time  for $\chi^{-},\tau_{12}$.  Eq.~(\ref{Eq:ODE1}) may then be used to deduce that $\partial u/\partial y$ is also a function only of $t$. Combined with the boundary conditions of Eq.\eqref{bcic} this yields
  \begin{eqnarray*}
      \frac{\partial u}{\partial y} = U(t). 
  \end{eqnarray*}
Hence Eqs.\,\eqref{gend40},\eqref{gend41} reduce to the ordinary differential equations 
\begin{eqnarray}\nonumber 
  \lambda   \frac{{\rm d} \chi^{-}}{{\rm d}t} &=& - \chi^{-} +  \frac \alpha \delta U(t) \tau_{12}, \\
     \lambda  \frac{{\rm d} \tau_{12}}{{\rm d} t} &=& -\tau_{12} + \frac 1 {\delta } U(t)( 1- \lambda \chi^{-}),\label{odes}
\end{eqnarray}
where the slip coefficient is in the range  $a\in [-1,1]$ and $\alpha= \lambda(1-a^{2})\geq 0$.

\noindent 
\subsection{The Oldroyd and Newtonian limits.}\label{ulnls}

\noindent 
In the Oldroyd limits we have $a\pm 1,$ so that $a^2=1,~\alpha=0$, and zero stress initial conditions so that both  $\chi^-,\tau_{12}$  are zero initially.  We then have 
$\chi^+=\chi^- = 0$ for all time.  From the definitions of Eqs.\,\eqref{chidef}, this entails that $\tau_{22}=0$ and thus from  the second equation of \cref{gend0}, we have $\partial p/\partial y =0$ and thus the pressure satisfies $p=p_*(t)$.  Hence the pressure is, at most, a function of time and does not impact on the fluid flow and but instead  simply balances the normal load across the joint. 

With $\chi^-=0$, we also have that solving for $ \tau_{12}$ gives 
\begin{equation}\label{ulconvect} \tau_{12}=  \dfrac 1 {\lambda \delta} \int_0^t \mathrm{e}^{-(t-s)/\lambda}U(s)\mathrm{d}s, 
\end{equation} 
and with $U(t)=\sin(2\pi \nu t)$, this collapses to 
\begin{equation}\label{ulconvect1} \tau_{12}=
\frac 1 \delta \dfrac{1}{1+4 \lambda^2\nu^2 \pi^2} \left(\sin(2\pi\nu t)-2\lambda \nu\pi\cos(2\pi\nu t) +2\lambda \nu\pi \mathrm{e}^{-t/\lambda}\right).
\end{equation} 
Hence,  the tangential stress scales like $1/\delta$, analogously to the Newtonian case, and since $\delta \ll 1$ extremely  high oscillatory tangential stresses are predicted within the synovial layer  with oscillatory joint motion. 

More generally  note that if $2\pi \nu \lambda \ll 1$, corresponding to a sufficiently small non-dimensional relaxation time, so that initial transients are very rapid, the solution with $U(t)=\sin(2\pi \nu t)$, as given by Eq.\,\eqref{ulconvect1}, satisfies 
\begin{equation}\label{ulconvect2} \tau_{12}\approx
\frac 1 \delta \sin(2\pi\nu t) = \frac 1 \delta U(t) =  \frac  1 \delta \frac{\partial u}{\partial y} .
\end{equation} 
This is the behaviour of a Newtonian fluid. Such Newtonian behaviour follows  from Eq.\,\eqref{ulconvect} for more general relative motion of the surfaces, given sufficiently small $\lambda$, via a standard application of  Watson's Lemma \citep{bender1999}.  
Instead if we have $a^2\neq 1$ but the
non-dimensional relaxation time is small enough to ensure 
$\lambda \ll \delta/(1-a^2)$ then $\alpha/\delta \ll 1 $ so that from Eqs.\,\eqref{odes} we have 
$$  \lambda   \frac{{\rm d} \chi^{-}}{{\rm d}t} \approx - \chi^{-}  , $$ and  thus $\chi^- \approx 0$ with the zero stress boundary conditions. Hence, 
Eq.\,\eqref{ulconvect} is an approximate solution, as is  \eqref{ulconvect1} for $U(t)=\sin(2\pi \nu t)$, so that the Newtonian limit emerges, as expected, for sufficiently small relaxation times. However, even for the upper and lower convected fluids $\nu \lambda = \nu^* T^* > 2$ and hence the Newtonian limits are not pertinent to joint motion, except for confirming that there is a consistent Newtonian limit. 

\subsection{Other constitutive relations, steady motion}
Below we assume $\alpha \gg \delta^{2/3}$, noting the power of $2/3$ in this relation is deduced \textit{a posteriori}. With this constraint in place,  we are not dealing with a viscoelastic fluid that has either an upper or lower convected derivative, nor close to one on a scale measured by $\delta^{2/3} \ll 1.$ Thus, the common Jaumann derivative among others is considered.

It is instructive to consider $U(t)=U,$ constant in the first instance, even though this is not pertinent for a joint as a relative constant  motion, without a reversal of direction,  is not feasible. Then, with $J,~K$ constants that are determined from the zero stress initial conditions, we have that the two coupled autonomous linear ordinary differential equations of \eqref{odes}  are solved by 
\begin{eqnarray}
 &  \tau_{12}=  \dfrac{\delta U}{\delta^2+\alpha\lambda U^2}
 +  \mathrm{e}^{-t/\lambda}\left[J\sin\left( \sqrt{1-a^2} \dfrac{Ut}\delta \right)+K\cos\left( \sqrt{1-a^2} \dfrac{Ut}\delta \right) \right] ,& \nonumber \\ 
& 1- \lambda \chi^- = \dfrac{\delta}U (\lambda\dot{\tau}_{12}+\tau_{12}). &\label{static0} 
\end{eqnarray}
Hence we see that the fast timescale of $t/\delta$ only occurs in the oscillations and the relaxation to the steady solution 
\begin{equation} \frac{\delta U}{\delta^2+\alpha\lambda U^2} \label{static} 
\end{equation}
is on the slow timescale. Thus there is no fast relaxation and the steady state solution may not be reached and thus may not be fully informative.  Nonetheless, noting the solution has a completely different  scaling compared to the upper or lower convected derivatives, with the shear  stress, $\tau_{12}$,  scaling  with $\delta$, in turn highlights that dramatically lower tangential stress may arise with different viscoelastic fluids. Hence, the  tangential stress in a joint may be extremely sensitive to the properties of synovial fluid. 

\subsection{Other constitutive relations, time varying.}
Once more we assume $\alpha \gg \delta^{2/3}$, so that we are not dealing with a viscoelastic fluid that is based  on an invariant derivative that is upper or lower convected, nor is close to one and we consider  oscillatory joint motion. With the definitions 
$$q:=1-\lambda \chi^-, ~~~~ \bm q := (
q , \tau_{12})^{\top}$$ and eliminating $\chi^-$ for $q$, we have 
\begin{eqnarray}
    \frac{{\rm d} \bm q}{{\rm d}t} &=&
  \left( \begin{array}{c} 
1 /\lambda\\0 \end{array}\right)- \bm M(t) \bm q, ~~~~~~~
\bm M(t) :=   \dfrac 1 \lambda \left(
\begin{array}{cc} 
1 & \beta U(t)/\delta  \\ - U(t)/\delta & 1  \end{array}\right), ~~~\label{ges}
\end{eqnarray}
where 
\begin{equation} \label{beta} \beta= \alpha\lambda=\lambda^2(1-a^2)>0.
\end{equation}
We also note that commutation for $\bm M$ is trivial
for all $s,t$, that is 
$$ \bm M(t)\bm M(s)-\bm M(s)\bm M(t)=\bm 0.$$  Thus the usual complexities of matrix exponentiation, due to a lack of commutativity,  are not present and we have extensive simplifications of the path ordered exponential, also referred to as a time ordered exponential,  and its inverse: 
\begin{eqnarray*}
P\exp\bm M(t) &:=& \tensor{I} +\sum_{n=1}^\infty \int_0^t \mathrm{d}t_1 \int_0^{t_1} \mathrm{d}t_2 \ldots \int_0^{t_{n-1}} \mathrm{d}t_n   \bm M(t_1) \bm M(t_2)\ldots \bm M(t_n) \\
&=&  \exp\left[\int_0^t \hspace*{-1mm} \mathrm{d}s \bm M(s) \right] = \left[\exp\left[-\hspace*{-1mm} \int_0^t \hspace*{-1mm} \mathrm{d}s \bm M(s) \right]\right]^{-1} \hspace*{-1.5mm} = [ P\exp [- \bm M(t)]]^{-1}.
\end{eqnarray*} 
In particular, commutation means these relations are immediately  inherited from results for the exponentiation of scalars. With 
$\bm q_0 = \bm q(t=0)$ the initial condition, the solution of Eq.\,\eqref{ges} is given by 
\begin{eqnarray*}
\bm q &=& P\exp[-\bm  M(t)] \bm q_0 + P\exp[-\bm M(t)]\int_0^t P\exp\bm M(s ) \mathrm{d}s 
  \left( \begin{array}{c} 
1 /\lambda \\0 \end{array}\right)
\\ 
&=& 
\exp\left[-\int_0^t \hspace*{-1mm} \mathrm{d}s \bm M(s) \right] \bm q_0 +
\int_0^t \exp\left[-\int_s^t\mathrm{d}r \bm M(r) \right]  \mathrm{d}s 
 \left( \begin{array}{c} 
1 /\lambda \\0 \end{array}\right).
\end{eqnarray*} 
Noting 
$$ \int_0^t  \mathrm{d}s \bm M(s)  = \dfrac t \lambda \tensor{I} + \frac 1 {\lambda \delta}  \int_0^t U(s) \mathrm{d}s
\left(
\begin{array}{cc} 
0 & \beta  \\ -1 & 0  \end{array}\right),
$$
we have 
\begin{eqnarray*} \exp\bigg[-\int_s^t \hspace*{-1mm} \mathrm{d}r  && \bm  M(r) \bigg]  =  \\   e^{-(t-s)/\lambda} &&  
\left(
\begin{array}{cc} 
\cos\left(\dfrac{1} {\delta\lambda} \sqrt{\beta} \int_s^t  \mathrm{d}q ~U(q) 
\right) &  -\sqrt{\beta} \sin\left(\dfrac{1} {\delta\lambda} \sqrt{\beta} \int_s^t  \mathrm{d}q ~U(q) 
\right)  \\ \dfrac 1 {\sqrt{\beta} } \sin\left(\dfrac{1} {\delta\lambda} \sqrt{\beta} \int_s^t  \mathrm{d}q ~U(q) 
\right) & \cos\left(\dfrac{1} {\delta\lambda} \sqrt{\beta} \int_s^t  \mathrm{d}r ~U(r) 
\right)  \end{array}\right).
\end{eqnarray*} 
 In particular for zero initial stresses, so that $q(t=0)=1, \tau_{12}(t=0)=0$, yielding $\bm q_0 = (1,0)^\top$, we have  the transient contribution to the stresses in the above is given by 
\begin{eqnarray}\label{fapt}
q^{\mbox{{\small trans}}} = e^{-t/\lambda}  \cos\left(\dfrac{1}{\delta\lambda}  V(t)\right) , ~~~~~~~
\sqrt{\beta} \tau_{12}^{\mbox{{\small trans}}} = e^{-t/\lambda}  \sin\left(\dfrac{1}{\delta\lambda}  V(t)\right)  . ~~~~~~~
\end{eqnarray}
where 
$$  V(t) = \sqrt{\beta} \int_0^t  \mathrm{d}r ~U(r) .$$ In turn this may be summarised as 
\begin{equation}\label{trans} q^{\mbox{{\small trans}}} +\mathrm{i} 
\sqrt{\beta} \tau_{12}^{\mbox{{\small trans}}} = e^{-t/\lambda}  \exp\left(\dfrac{\mathrm{i}}{\delta\lambda}  V(t)\right)  ,
\end{equation}
with $q^{\mbox{{\small trans}}} ,~ \sqrt{\beta}  \tau_{12}^{\mbox{{\small trans}}}  $ given via the real and imaginary parts of the right-hand-side. 

Furthermore, with $W(r):=V(\lambda\ln (r/\lambda))$,  we have the stresses are governed by 
\begin{eqnarray} \label{fapg}
q +\mathrm{i} 
\sqrt{\beta} \tau_{12} &=&   \mathrm{e}^{-t/\lambda} \exp\left[  \dfrac{\mathrm{i}V(t)}{\delta\lambda} \right]\left[1+ \frac 1 \lambda
\int_0^t  \mathrm{d}s  \exp\left[\dfrac{s}{\lambda}- \dfrac{\mathrm{i}}{\delta\lambda}  V(s)\right]
\right] \\  \nonumber  &=& 
\mathrm{e}^{-t/\lambda} \exp\left[  \dfrac{\mathrm{i}V(t)}{\delta\lambda} \right]\left[1+ \frac 1 \lambda
\int_1^{\lambda \mathrm{e}^{t/\lambda}}  \mathrm{d}r \exp\left[- \dfrac{\mathrm{i}}{\delta\lambda} W(r)\right]\right].
\end{eqnarray}
Note that the integral term is not a transient as it scales like exp($t/\lambda$) for large time and thus does not simply decay due to the exp($-t/\lambda$) pre-factor. 
In addition, the final expression is written in a form that allows the direct application of the stationary phase approximation \citep{bender1999}, with $1/(\lambda\delta)$ the large parameter. Assuming that the time, $t$, is sufficiently large to ensure that:
\begin{itemize} 
\item [(i)] the interval $[0,t]$ contains a root of $U(t_*)=0$, with all such roots  single, that is $U'(t_*)\neq 0$; and 
\item[(ii)] the exponential decay of $\tau_{12}^{\mbox{{\small trans}}}$ renders it negligible, 
\end{itemize} 
we have the stationary phase approximation 
\begin{eqnarray} & q+\mathrm{i} \sqrt{\beta} \tau_{12} \approx
\dfrac{\sqrt{2\pi\delta}}{\sqrt{\lambda}\beta^{1/4}} 
\sum_{t_*} \dfrac{e^{-(t-t_*)/\lambda} }{|U'(t_*)|^{1/2}}
\exp\left[ 
-\mathrm{i}\dfrac{\pi}{4}\mbox{sgn}(U'(t_*))+\dfrac{\mathrm{i}\sqrt{\beta}}{\delta\lambda} \int^t_{t_*}  \mathrm{d}r \, U(r) \right]  , &  \nonumber \\   &
t_*\in \left\{r  \Big| U(r)=0,~~ 0\leq r <t \right\}.& \label{stphap}
  \end{eqnarray} 
However, this stationary phase approximation requires that the domain around the roots of $U$  can be extended to infinity, requiring 
$$ |t-t_*| \gg \lambda^{1/2} \delta^{1/2},$$
with a concomitant loss of accuracy once $t$ is sufficiently close to a root of $U$. However, the loss of accuracy is localised and due to the limitations of the stationary phase method rather than large, local changes in either the approximation or the behaviour of the solutions. Hence, the stationary phase approximation will not be in gross error in these localised regions, even if precision is lost. 
  
  Furthermore, with 
  $$t_*^{max}(t) := \mbox{max} \left\{t_* \Big| U(t_*)=0,~~ t_* < t \right\}$$ defined to the largest root of $U(t_*)=0$ that does not surpass $t$ and provided the roots of  $U(t_*)=0$ are sufficiently separated in time to ensure that 
 $$ e^{t_*^{max}/\lambda} \gg e^{t_*/\lambda}, ~~~~ \mbox{for all} ~t_*\in \left\{t_* \Big| U(t_*)=0,~~ t_* < t_*^{max} \right\}, $$ this simplifies to 
\begin{eqnarray}\nonumber 
q +\mathrm{i} 
\sqrt{\beta} \tau_{12} &\approx& \dfrac{\sqrt{2\pi\delta}}{\sqrt{\lambda}\beta^{1/4}}   \dfrac{e^{-(t-t^{max}_*)/\lambda} }{|U'(t^{max}_*)|^{1/2}}
\exp\left[  -\mathrm{i}\dfrac{\pi}{4}\mbox{sgn}(U^\prime(t^{max}_*))+
\dfrac{\mathrm{i}\sqrt{\beta}}{\delta\lambda}  \int^t_{t^{max}_*}  \hspace*{-1.5mm} \mathrm{d}r   U(r) \right]  , \\ \label{stphap1} 
  \end{eqnarray} 
  again subject to $ |t-t^{max}_*| \gg \lambda^{1/2}\delta^{1/2},$ 
which will hold for most times since $\delta^{1/2}\ll 1$.

Thus, after the relaxation of transients, the tangential  stress $(\tau_{12})$ under relatively weak conditions for the forcing velocity $U(t)$ is an oscillating function with a small maximum amplitude of 
\begin{equation} {\cal A}_\tau= \dfrac{\sqrt{2\pi \delta}}{\sqrt{\lambda} \beta^{3/4}|U'(t^{max}_*)|^{1/2}} = \dfrac{\sqrt{2\pi \delta}}{\lambda^2 (1-a^2)^{3/4}|U'(t^{max}_*)|^{1/2}}  \sim \mbox{ord}(\delta^{1/2}), \label{scaling} 
\end{equation} 
and a oscillation phase dependence that varies in time and with the details of the forcing $U(t)$. Note that for $1-a^2\gg\delta^{2/3}$ we have that ${\cal A}_{\tau}$ is small given all other parameters are order unity. 
Similarly the magnitude of $q=1-\lambda \chi^-$ in its oscillations is given by 
\begin{equation}  \sqrt{\beta}{\cal A}_q = \dfrac{\sqrt{2\pi \delta}}{\lambda (1-a^2)^{1/4}|U'(t^{max}_*)|^{1/2}}  \sim \mbox{ord}(\delta^{1/2}) \label{tau22}
\end{equation}
so that 
$ \chi^-= 1/\lambda + O(\delta^{1/2})$.

Finally, recalling $\chi^+=0$ and the definitions of Eqs.\,\eqref{chidef}, we have $$\chi^{-} = -(1+a)\tau_{22}$$ {\mg to give $\tau_{22}$, the normal stress, and from the second of Eqs.\,\eqref{gend0} we have the pressure satisfies  $p=p_*(t)$, analogously to the derivation of this observation for the Oldroyd models.} Hence, once more, 
 the pressure  simply balances the normal load across the joint. 

\subsection{The Stationary phase approximation for harmonic motion}
We consider a harmonic driving motion with
$$ U(t) = \sin(2 \pi\nu t) , $$
of unit maximum magnitude, by the choice of the non-dimensionalisation scale, $W^*$, and the resulting simplification of the stationary phase approximation given by 
Eq.~\eqref{stphap}, given that transients have decayed. With the derivation presented in the  Appendix  and  the definitions 
\begin{eqnarray*}
s_1 &:=&  \sin   \left( \dfrac{(1-a^2)^{1/2}}{2\pi\nu\delta}(1-\cos(2\pi\nu t))-\dfrac \pi 4  \right) \\
s_2 &:=&  \sin\left( \dfrac{(1-a^2)^{1/2}}{2\pi\nu\delta}(1+\cos(2\pi\nu t))-\dfrac \pi 4  \right)
\end{eqnarray*}
the tangential stress on the $N^{th}$ oscillation, with $N$ even, is approximated by 
\begin{eqnarray}\label{fap1m}
\tau_{12} &\approx& 
\dfrac{\delta^{1/2} \e^{-(t-t_*^{max})/\lambda}} {\lambda^{2}\nu^{1/2}(1-a^2)^{3/4}(1-\e^{-1/(\nu\lambda)})} \left(s_1 - \e^{-1/(2\nu\lambda)}s_2\right) . 
\end{eqnarray}
Analogously,  for $N$ odd, the approximation is 
\begin{eqnarray}\label{fap2m}
\tau_{12} &\approx& 
\dfrac{\delta^{1/2} \e^{-(t-t_*^{max})/\lambda}} {\lambda^{2}\nu^{1/2}(1-a^2)^{3/4}(1-\e^{-1/(\nu\lambda)})} \left(  \e^{-1/(2\nu\lambda)}s_1-s_2\right) .
\end{eqnarray}
Note that as  $\nu t$ passes through a half-integer value there is a loss of accuracy, and continuity, of this stationary phase approximation. Away from such points, it does generate a good approximation, especially for smaller $\delta$, as may be observed in the results of Fig.\, \ref{fig2} and demonstrates the complex dependence predicted between the physical parameters and the shear  stress.  {\bl In particular one can observe from the structure of the sinusoidal terms in Eqs.~(\ref{fap1m}),(\ref{fap2m}) that the tangential stress will oscillate in time with a frequency that will scale with 
\begin{eqnarray}\label{freq}
 \dfrac{(1-a^2)^{1/2}}{2\pi\nu\delta},
\end{eqnarray}
and hence high temporal frequencies are expected since $\delta \ll 1.$ 
}

 \section{Numerical Validation}\label{sec5} 

\begin{figure}
 \begin{center}
  \hspace*{-1.3cm} \includegraphics[width=15.8cm]{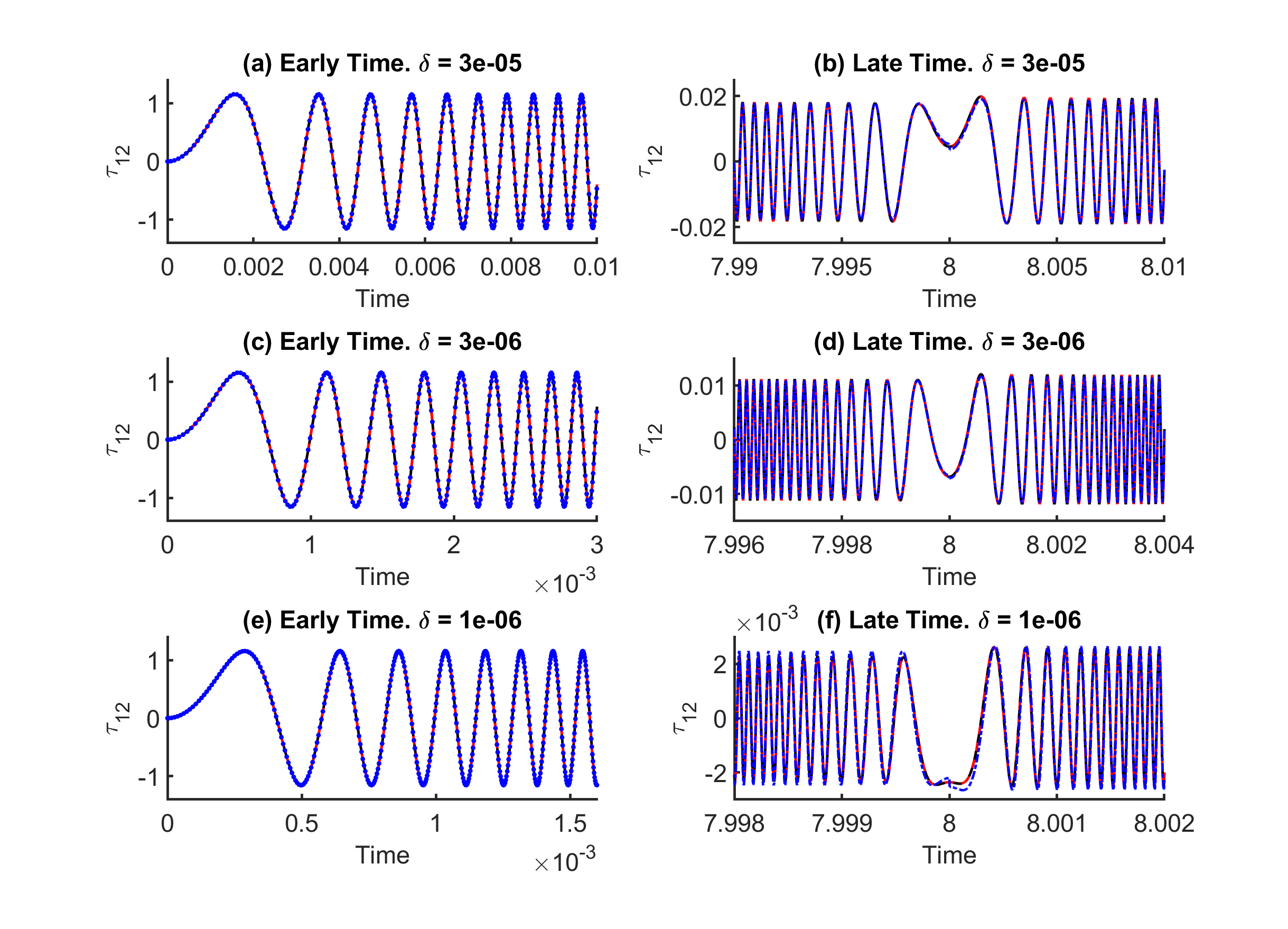}
 \end{center} 
 \caption{A comparison of the  approximation of 
 Eqs.\,\eqref{fapg},~\eqref{fap1m},~\eqref{fap2m} to the solutions of the ordinary differential equations, Eqs.\,\eqref{odes}, for the parameter values 
 $\nu = 3.5, ~a=0.5, ~\lambda = 1.0, ~\beta=3/4 $, with the non-dimensional channel thickness given by the value of $\delta$ specified in the plots. {\bf (a)(c)(f)} Early time plots, where the red dash curve is the numerical solution for  Eqs.~\eqref{odes};  the analytical  solution of Eq.\,\eqref{fapg} is also plotted in black dash, but is difficult to observe in the presence of the   blue open circles as the numerical and analytical are effectively the same at this plotting resolution. The blue open circles correspond to the transient solution for $\tau_{12}$, from Eq.\,\eqref{trans} and simplifying to Eq.\,\eqref{shst} and also agree with the numerical and full analytical solutions at these times to within plotting  resolution. {\bf (b)(d)(f)} Late time plots. Once more the numerical solution is presented (red dash) with the  analytical solution (black dash); these  effectively agree to within plotting resolution, making the distinction of the red and black dash difficult to observe. The  stationary phase approximation given by Eqs.~\eqref{fap1m},\eqref{fap2m} is presented in blue dash and gives excellent agreement, though with a mild loss of accuracy at zeros of $U(t)$, one of which occurs  in these plots at $t=8.0$.}
 \label{fig2}
\end{figure}

We proceed to confirm observations from the analytical structure of the solutions to Eqs.\,\eqref{odes} with zero stress initial conditions and a driving velocity of $U(t)=\sin(2\pi\nu  t),$ with $ \nu=3.5$, a typical non-dimensional frequency as noted in Table \ref{tab2}. In particular, in Fig. \ref{fig2} we compare solutions of these equations for the tangential stress, $\tau_{12}$ to the analytical solution of Eq.\,\eqref{fapg} and the stationary phase approximations of Eqs.\,\eqref{fap1m},\eqref{fap2m} for representative parameter values.

\begin{figure}
 \begin{center}
 \hspace*{-0.7cm} \includegraphics[width=14.4cm]{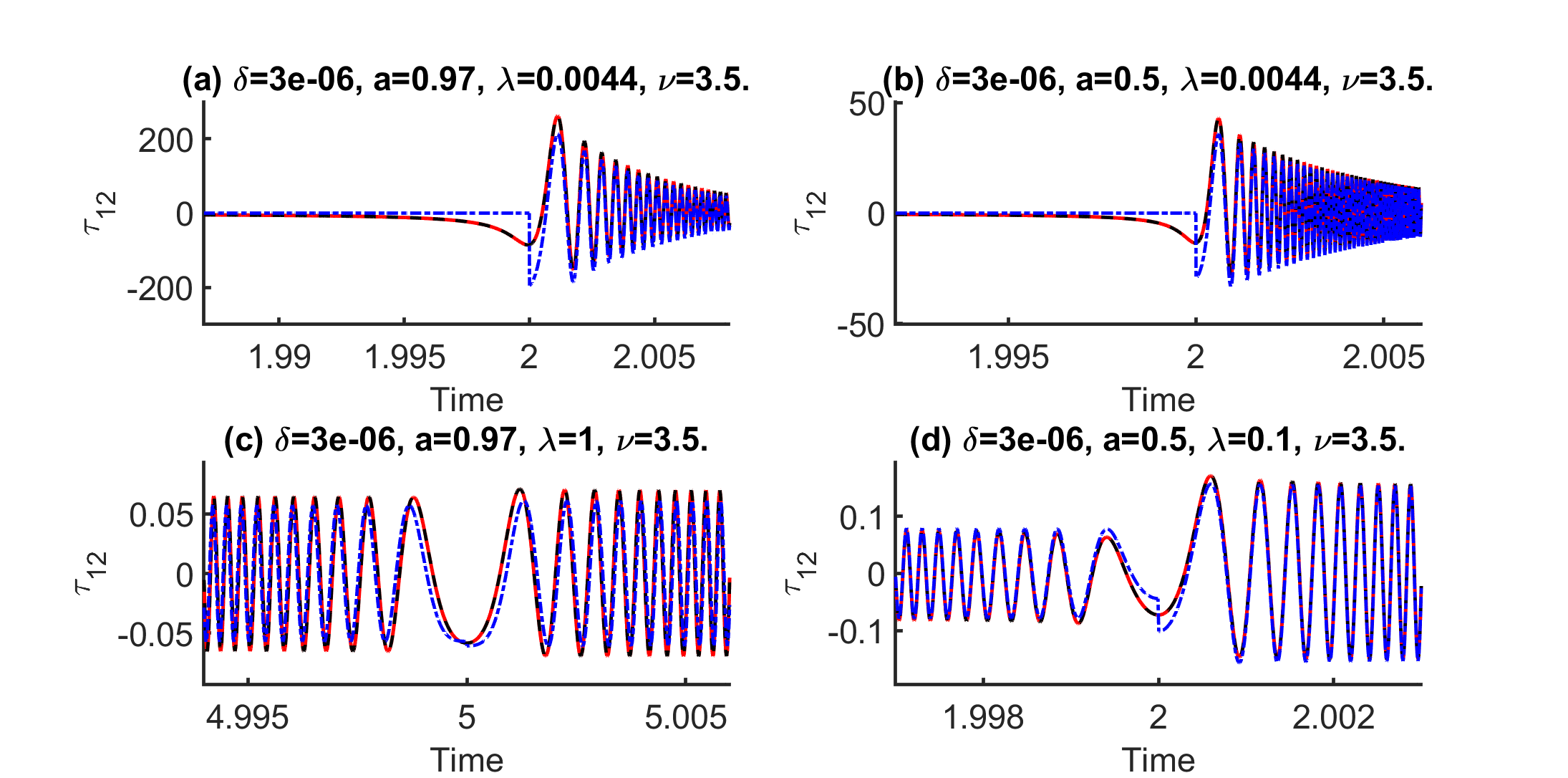}
 \end{center} 
 \caption{Plots of the  at late time, after transients have decayed, for extremes of parameters. In plots (a),(b)  $\lambda=0.02s/T^*\sim 4.4\times10^{-3}$, corresponding to a loss of elasticity, at the level of disease for most estimates in Table \ref{Tab:Pars}. With the estimate of $a$ from \citet{lai1978}, there is a clear increase in the overall scale of the tangential stress by noting the vertical axis. However, changing the slip parameter from $a=0.97$ to $a=0.5$ reduces the stress, as seen by comparing (a), (b)  and thus even polymer prolateness can reduce predictions of friction.  as can also be observed in comparing plot (c) with Fig\ref{fig2}(d). Similarly comparing Fig\ref{fig2}(d) to plot (d) highlights that reducing elasticity, that is $\lambda$, increases stress.
 }
 \label{fig3}
\end{figure}

\subsection{Accuracy of the mathematical approximation, Eqs.~(\ref{fap1m}),(\ref{fap2m})}
One can observe from all plots that the analytical expression of 
Eq.\,\eqref{fapg}, as presented by the black-dash curves, agrees  to within plotting resolution of the numerical solution for the differential equations Eqs.\,\eqref{odes}, as given by red-dash curves. The agreement is to the extent that it is very difficult to observe the black-dash.  
One can also observe the dramatic increase in the frequency of the  with reducing $\delta$, even though the frequency of the forcing is fixed. There is also 
 complex behaviour, as expected, around the zeros of the boundary motion velocity $U(t)=\sin(2\pi\nu t)$, as seen here in plots (b), (d), (f) around $t=8.0$.  
For the early time plots, as presented in plots (a), (c), (e) of Fig. \ref{fig2} we also present the transient for $\tau_{12}$ via blue open circles as given by 
\begin{equation}\label{shst} \frac 1 {\sqrt{\beta}} e^{-t/\lambda}  \sin\left(\dfrac{\sqrt{\beta}}{\delta\lambda}  \int_0^t  \mathrm{d}r ~U(r)\right)  =  \frac 1 {\sqrt{\beta}} e^{-t/\lambda}  \sin\left(\dfrac{\sqrt{\beta}}{2\pi\nu\delta\lambda}  \left[1-\cos(2\pi\nu t )\right]\right) 
\end{equation}
from Eq.~(\ref{trans}). {\mg As expected the transient dominates for $t\ll \lambda$ and thus at early time we have that the non-dimensional tangential stress is predicted to have a maximum magnitude of the order of $O(1/\sqrt{\beta})$ for the initial stages of joint motion. }

For the late time plots,  after the decay of transients, as presented in plots (b), (d), (f) of Fig. \ref{fig2} we also present the stationary phase approximations of Eqs.\,\eqref{fap1m},\eqref{fap2m}  for $\tau_{12}$ via the blue curves. These give an excellent approximation, though with the loss of accuracy in the vicinity of the roots of $U(t)$ due to the limitations of the asymptotic method. Nonetheless, the loss of accuracy is localised and small, so that   the stationary phase still gives a suitable estimate in these regions. In particular, the loss of accuracy is not due to a singular behaviour of the solution but the inability to extend the domain of integration to infinity after making asymptotic approximations, a core feature of many integral asymptotic methods.  
The reduction in size of  oscillation amplitude after initial transients have decayed can also be  observed as $\delta$ decreases by examining plots  (b), (d), (f); {\mg more generally, the small magnitude of the tangential stress in these oscillations is particularly noteworthy.}

\subsection{Changes in the  on varying elasticity parameters and fluid depth}\label{sec52} 
We proceed to consider the behaviour of the approximation and the numerical solutions at more extremes of the parameter space, focussing on late time solutions, after initial transients have decayed. 
In Figs.~\ref{fig3}(a),(b) we have reduced the elasticity to values typically seen in disease with $\lambda^*=0.02$ seconds, $\lambda = 4.4\times 10^{-3}$ (see Tables \ref{Tab:Pars},\ref{tab2}), and in plot (a) we have also increased the slip parameter to  $a=0.97$, the value indicated by \citet{lai1978}. One can immediately see that Eqs.~(\ref{fap1m}),(\ref{fap2m})  maintain accuracy  into pathophysiological regimes and 
that physiological elasticity reduces tangential stress, and thus friction. In particular,  the predicted tangential stress  after transients has increased by more than three orders of magnitude with the change in elasticity: compare Fig.~\ref{fig2}(d) with Fig.~\ref{fig3}(b), with the sensitivity arising from the observation that $\tau_{12}\sim O(1/\lambda^2)$ in Eqs.~(\ref{fap1m}),(\ref{fap2m}). Increasing the prolateness of the polymer inducing elasticity also increases friction, as observed by comparing Fig.~\ref{fig3}(a) with Fig.~\ref{fig3}(b),  Fig.~\ref{fig2}(d) with Fig.~\ref{fig3}(c), or by noting 
$$ \tau_{12} \sim \frac 1 {(1-a^2)^{3/4}}$$ 
in Eqs.~(\ref{fap1m}),(\ref{fap2m}). However, even with $a=0.97$ elasticity is still protective and prolateness is still influential, as the magnitude of the tangential stress seen in Fig.~\ref{fig3}c is still much less than that of the Newtonian fluid, or an Oldroyd-B fluid. For the latter the stress is given by \cref{ulconvect1} after initial transients, and thus (by use of double angle formulae and the parameters of Fig.~\ref{fig3}(c)) is of magnitude 
\begin{equation}\label{shrthin} \frac{1}{\delta} \frac 1  {\sqrt{1+4\lambda^2\nu^2\pi^2}} \sim  10^{4}.
\end{equation}
Hence, even though \citet{lai1978} estimated $a=0.97\approx 1$, and suggested the use of an Oldroyd-B fluid which corresponds to $a=1$, the singular limit of the stress as $a\rightarrow 1$ Eqs.~(\ref{fap1m},\ref{fap2m}) demonstrates that the prediction of stress is fundamentally changed on taking the Oldroyd-B limit, and thus this is not a good modelling choice. 


  \begin{figure}
  \begin{center}
 \hspace*{-0.3cm}  \includegraphics[width=14.5cm]{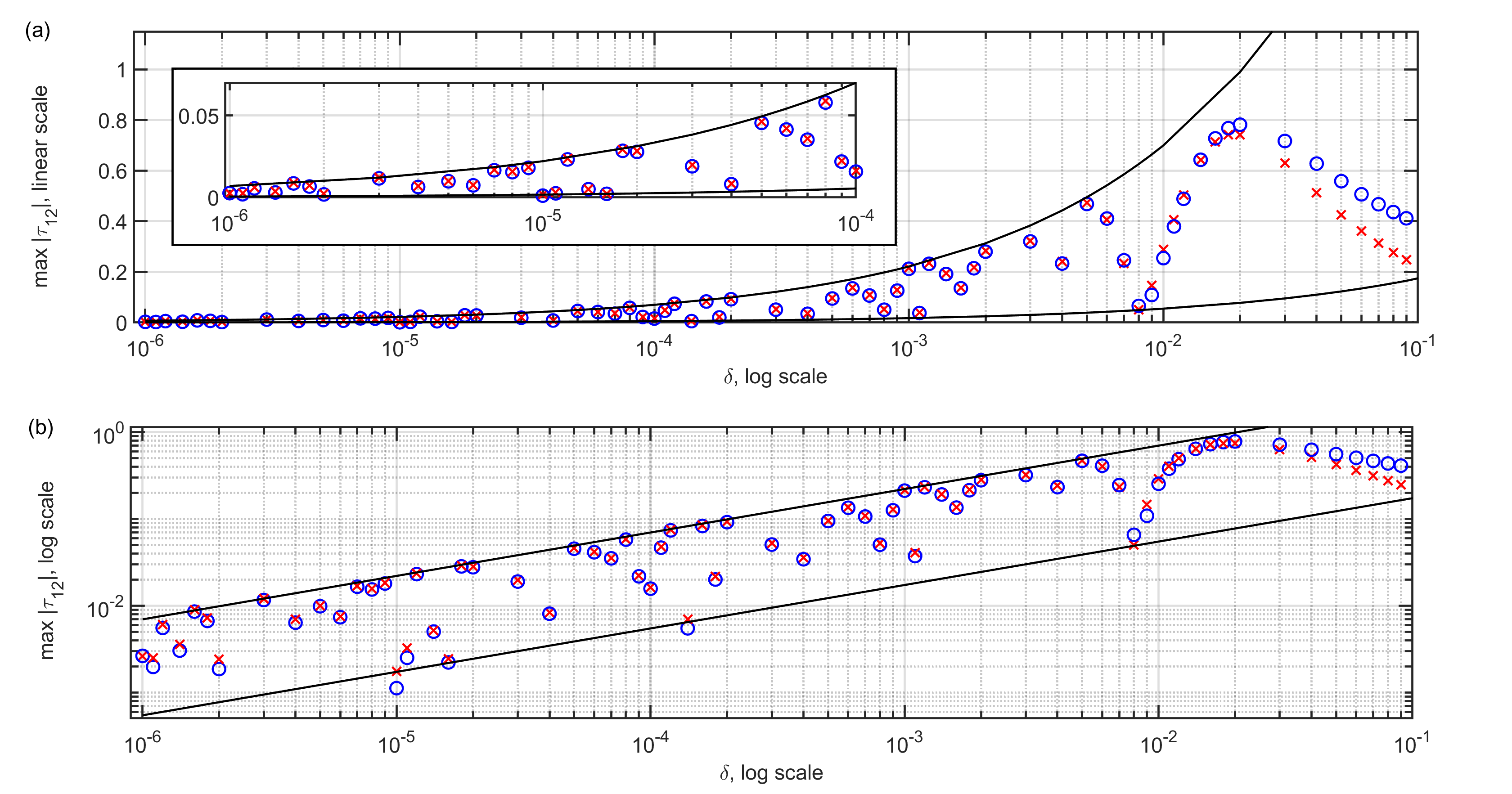}
  \end{center}
  \caption{The scaling behaviour of the maximum amplitude of $|\tau_{12}|$ after initial transients, for $a=0.5,\lambda=1,\nu=3.5$ with $\delta$ varying along the horizontal axes on a log scale. In both plots the red crosses are the results from numerical solution of the ordinary differential equations, Eq.~\eqref{odes}, and the blue crosses are the stationary phase approximations, Eqs.~(\ref{fap1m}),(\ref{fap2m}). The black curves are  plot of a constant multiple of $\delta^{1/2}$, and thus shows a $\delta^{1/2}$ scaling for peaks and troughs 
  of the variation in the maximum amplitude of $|\tau_{12}|$ as $\delta$ changes. 
  The upper plot, (a), presents the  maximum amplitude of $|\tau_{12}|$ on a linear scale, while the lower plot, (b), uses a log scale. The inset in plot (a) is a magnification at lower values of $\delta$ for clarity.}
  \label{Fig:ODE}
  \end{figure}
\begin{figure}
  \begin{center}
 \hspace*{-0.4cm}  \includegraphics[width=1.34\linewidth]{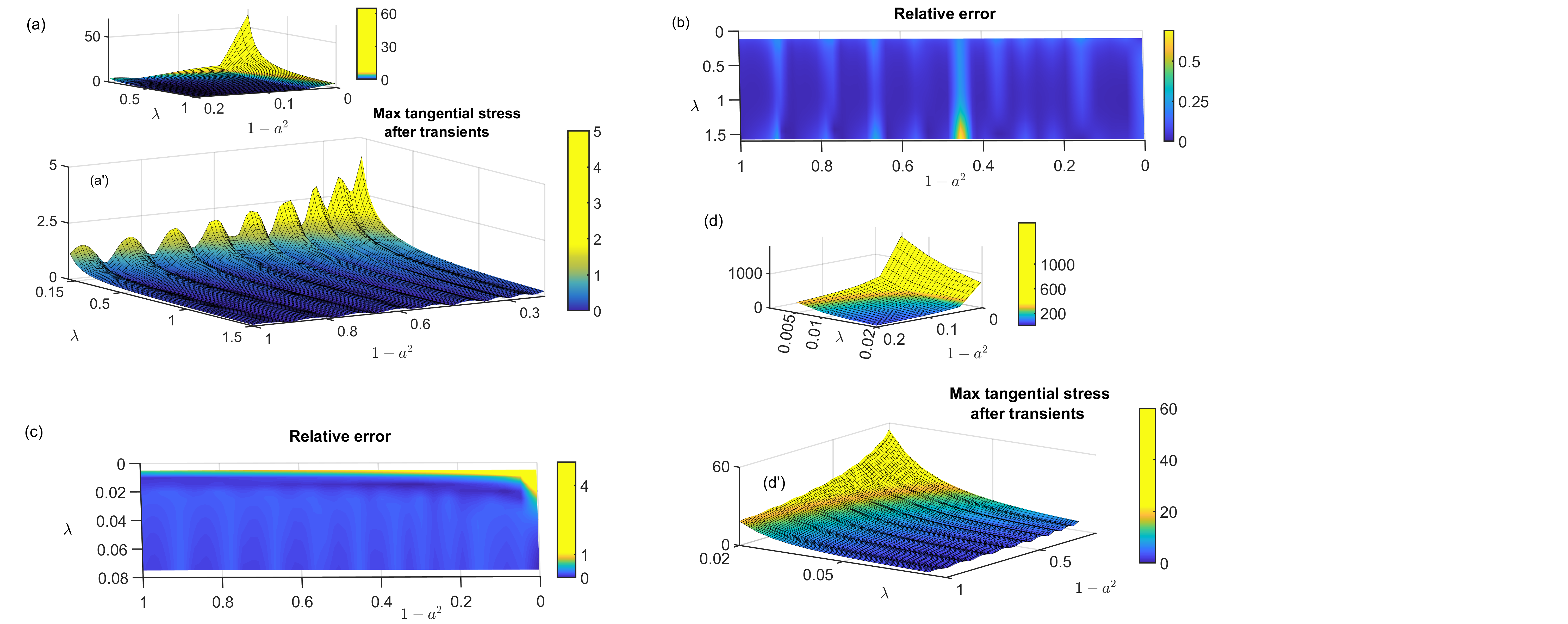}
  \end{center}
  \caption{In plots (a), (a$'$), (d), (d$'$) the scaling behaviour of the maximum amplitude of $|\tau_{12}|$ after initial transients is presented in terms of the non-dimensional elastic relaxation time, $\lambda$ and the slip parameter, $a$, noting the dependence on $a$ is given in terms of $(1-a^2)$ on the horizontal axes.  Here we have fixed $\delta = 5\times 10^{-4}, ~\nu = 3.5$.  In plots (a), (a$'$) we have $\lambda > 0.08$, {\mg with $\lambda<0.08$ in plot (d$'$). In plot (a) we also have $1-a^2 < 0.2$ and for plot (d), $1-a^2<0.2$, $\lambda < 0.02$. 
  The need to present different plots  for separate values} of the non-dimensional elastic relaxation time, $\lambda$ and the slip parameter, $a$, is because of the orders of magnitude variation in the maximum amplitude of $|\tau_{12}|$ after transients as $1-a^2$ and $\lambda$ approach zero, as expected from the scaling relations of Eqs.\,(\ref{scaling}),\,(\ref{fap1m}),\,(\ref{fap2m}). 
  One can observe the overall increase in the tangential stress after transients with   decreasing $(1-a^2)$ predicted in the scaling relation, with an  additional  oscillatory dependence since $(1-a^2)$ also contributes to the frequency, Eq.\,\eqref{freq}, within the  trigonometric functions of
  Eqs. (\ref{fap1m}),\,(\ref{fap2m}). In contrast
  decreasing $\lambda$, does not induce oscillation and simply results in an increase in the tangential stress, consistent with the dependencies give in 
 Eqs.\,(\ref{scaling}),\,(\ref{fap1m}),\,(\ref{fap2m}),\,(\ref{freq}). {\mg Thus, as the elasticity reduces, or the prolateness of the elasticity inducing polymer increases, there is an increase in the maximal tangential stress after transients,  extensively  so for extreme parameter values.}  In plots (b), (c) the relative error of the relations (\ref{fap1m}),\,(\ref{fap2m}) compared to the solutions of the differential equations (\ref{odes}) are presented, respectively for the range of $\lambda$ and  $a$ in plots (a,a$'$) and (d,d$'$). While in (b) the relative error can start to rise this occurs where the maximum amplitude of $|\tau_{12}|$  is still very small, so the absolute error remains very small. Hence for the parameters in the  plots (a,a$'$) the relations (\ref{fap1m}),(\ref{fap2m}) are accurate. However, the relative error becomes very large in plot (c) for small $\lambda,1-a^2$, where the maximum amplitude of $|\tau_{12}|$ is also extremely large; hence the approximations (\ref{fap1m}),\,(\ref{fap2m}) are no longer accurate in this region of parameter space.
  }
  \label{surface}
  \end{figure}

  We have also looked in a more refined manner at the link between the maximum tangential stress of $\tau_{12}$ once transients have decayed and the non-dimensional film thickness, $\delta$. In particular,  the simple scaling of 
$$ \tau_{12} \sim \delta^{1/2}$$ 
given by  Eqs.~(\ref{scaling}),(\ref{fap1m}),(\ref{fap2m}) 
has the potential to be perturbed by the fluid thickness dependence {\bl within the frequency, Eq.\,\eqref{freq}, of the sinusoidal functions in these equations.
This can  indeed be observed in the} semilog plot and the log-log plot of Fig.\,\ref{Fig:ODE}(a),(b) respectively, where the red cross is the numerical solution for the maximum of $|\tau_{12}|$ after initial transients and the blue circle is the result from Eqs.~(\ref{fap1m}),(\ref{fap2m}). Noting the black curves are constant multiples of $\delta^{1/2}$ in both plots this shows that the peaks and troughs of the variation in $ \tau_{12}$ 
still scale with $\delta^{1/2}$. Hence there is always an upper envelope that scales with $\delta^{1/2}$, so that an upper bound for  $ \tau_{12}$  has a simple scaling with $\delta^{1/2}$. Analogous comments also apply to the slip parameter, $a$, and the non-dimensional frequency, $\nu$  for the scalings 
$$ \tau_{12}\sim \frac 1 {(1-a^2)^{3/4}}, ~~~~~~~~~~~~ \tau_{12}\sim \frac 1 {\nu^{1/2}}, $$
as $a$, $\nu$ also appear {\bl within the frequency of the sinusoidal functions, as given by Eq.~(\ref{freq}).  This is also consistent with the predictions of how the maximum tangential stress varies with $1-a^2$ that are illustrated in Fig.\,\ref{surface}, which also shows that there are no oscillations as the non-dimensional elastic relaxation time, $\lambda$, is reduced. Instead, reducing $\lambda$ generates an inexorable rise in the maximal tangential stress after initial transients have decayed. 
In particular, within pathophysiology regimes of synovial fluid elasticity, which can correspond to {\mg $\lambda^* \sim 0.02\mbox{s},\,\lambda\sim 0.0044 $} from the parameter estimates of Tables \ref{Tab:Pars},\,\ref{tab2}, the rises are of multiple orders of magnitude, as observed in Figs.\,\ref{surface}(d,d$'$). Nonetheless the ratio of the shear stress, with elasticity, compared to Newtonian shear stress is given by 
\begin{equation}\label{rat}
\frac{\tau^{*}_{12}}{\tau^{*,Newtonian}_{12}} = \frac{\mu^*}{T^*}\tau_{12} \frac {d^*} {\mu^*W^*}= \delta \tau_{12}
\end{equation}
which is still small for the parameters {\mg associated with the levels of elasticity found in diseased joints, showing that elasticity is still predicted to  potentially ameliorate tangential stress even in disease. However,}
  this protection is ultimately lost as elasticity reduces ($\lambda\rightarrow 0$) or the magnitude of the slip parameter tends to unity ($a^2\rightarrow 1$).
}

More generally, the overlap of the numerical solution and the mathematical approximation in Fig.\,\eqref{Fig:ODE} until $\delta \sim 0.01$ shows that the  Eqs.~(\ref{fap1m}),(\ref{fap2m})
holds on increasing  the non-dimensional  fluid depth   beyond physiological values but such fluid depths are likely to still relevant for engineered lubrication mimicking a joint. Consequently, the impact of elasticity in reducing friction is also predicted to be inherited on scales more concomitant with engineering than joint physiology.


  \section{Discussion and Conclusions}\label{sec6} 

   Our objective  has been to examine how elasticity within models of synovial fluid mechanics impact predictions for tangential stress, and thus friction, for the length and timescales associated with oscillatory articular joint motion, and for lubrication more generally. 
     Here we have focused on differential constitutive relations, in particular considering a  Newtonian solvent and a viscoelastic polymer solute  with a Johnson-Segelman invariant derivative, 
   $$ \dfrac{D^{JS}\tau^*_{ij}}{Dt^*} := \dfrac{\partial \tau^*_{ij}}{\partial t^*} + u^*_p \tau^*_{ij,p} -\Omega^*_{ip}\tau^*_{pj}+\tau^*_{ip}\Omega^*_{pj}-a\left(\dot{{\cal E}}^*_{ip} \tau^*_{jp} + \tau^*_{ip}  \dot{{\cal E}}^*_{pj}  \right), ~~~ a\in[-1,1]. $$
Following a brief examination of numerical solutions {\mg for  small surface perturbations in \cref{micsc} to test robustness to such  perturbations,} we considered a one-dimensional model of shear between  flat parallel surfaces in relative motion. 
     While this is an extensive  simplification, it allows a study across the large plausible parameter space for a family of prospective models even with the uncertainties and variation in synovial fluid properties and the wide range of physical regimes in joint motion. 
     Furthermore, singular and sensitive behaviours, as observed here, are very unlikely to be removed by more complex geometries, or more complex constitutive  relations. In turn, this emphasises the importance of  elastic synovial fluid mechanics for understanding lubrication within the articular joint and, more generally, the potential for elastic contributions to lubricating fluid mechanics to ameliorate friction, whether in joints, artificial joints or more general  applications. 

   {\bl  For all the models considered, we observe  that the pressure is simply a function of time,  $p=p_*(t)$. Hence the pressure serves to balance loading, rather than contribute to the flow, as expected in that the fluid flow is being driven by joint motion rather than pressure gradients. However, the stress balances in the tangential directions are much more complex and vary between the models. For the canonical one-dimensional setting considered here,  one has that} the tangential stress in Newtonian fluids  will scale as   \begin{equation}\label{ntscal} \tau_{12}^{*,Newtonian} \sim\frac{\mu^*W^*}{d^*},
     \end{equation}
     where  $\mu^*, W^*, d^*$ are respectively the dimensional scales of the synovial fluid viscosity, the  maximal velocity of the joint motion and the fluid thickness. Hence, extremely  high tangential stresses emerge at small fluid thicknesses. 
   
   For the Oldroyd models, where the slip parameter satisfies $a=\pm 1$, we also observe from Eq.~\eqref{ulconvect} a scaling with $1/\delta$  for the non-dimensional tangential stress, which  on re-dimensionalisation, generates the scaling 
  \begin{equation}\label{obscal} \tau^{*,Oldroyd}_{12}\sim\frac{\mu^*W^*}{d^*} \frac 1{(1+4\lambda^{*2}\nu^{*2}\pi^2)^{1/2}} \sim \frac 1{2\pi} \frac{\mu^*W^*}{d^*\lambda^*\nu^*}  
  \end{equation}
   from Eq.\eqref{shrthin}, with the final relation holding for typical scales of elastic relaxation time, $\lambda^*$ and motion frequency, $\nu^*$. Hence there is once more very high tangential stresses, and thus friction, with decreasing fluid thickness. In addition, there is  shear-thinning with the peak stresses reducing as the shear rate, $\nu^*$, increases.

   The Newtonian scalings will emerge more generally for sufficiently small viscoelastic relaxation times, as noted after Eq.~\eqref{gend32}, and similarly the Oldroyd-B scalings will emerge for the slip parameter in the Johnson-Segelman derivative $a$ sufficiently close to unity. However, 
   neither the  Newtonian nor the Oldroyd limits have been observed to apply for the scales of joint motion, as noted in Section \ref{sec52} and evidenced in Fig.~\ref{fig3}. This figure also illustrates that  even the small levels of synovial fluid elasticity in rheumatoid arthritis are predicted to induce  reductions in friction.   
   This is consistent with the common understanding that the rheology of synovial fluid is important for its mechanics \citep{thurston1978,fam2007}. Furthermore,  noting the interpretation of $a$ in in terms of polymer prolateness \citep{hinch2021}, we can also observe that extreme prolateness is required for the Oldroyd-B limit. In particular, given the  non-dimensional fluid depth satisfies $\delta  \ll 1$  the mathematical expressions of Eq.~\eqref{scaling} show that  
   \begin{equation}\label{scob} (1-a^2) \gg \delta^{2/3}
   \end{equation}
   is sufficient to avoid the Oldroyd limits, as explicitly observed in Fig.~\ref{fig3}. Even  \citeauthor{lai1978}'s \citeyear{lai1978} estimate of $a=0.97$  satisfies Eq.~\eqref{scob} for the range non-dimensional fluids depth relevant for the joint, as given in Table \ref{tab2}.

With  $\delta  \ll 1, ~(1-a^2) \gg \delta^{2/3}$
the tangential stress in the constant shearing  case, as summarised by  Eq.\,\eqref{static}, indicates a radically different scaling. It also highlights that shear-dependent viscosity will not alter the fundamentally different scalings in that at constant shear the scaling is still very different once $$\delta  \ll 1, ~(1-a^2) \gg \delta^{2/3}.$$ 
However, other than these observations, we  find the static relation does not inform intuition for the oscillatory case representative of  joint motion time and lengthscales, where a very different relationship is observed. 

In particular,  after the decay of initial transients the non-dimensional tangential stress scales as (Eqs.\,\eqref{scaling}, \eqref{fap1m}, \eqref{fap2m}), 
\begin{equation} \label{fintau1} \tau_{12} \sim \dfrac{\delta^{1/2}} {\lambda^{2}\nu^{1/2}(1-a^2)^{3/4}} .
\end{equation}
By redimensionalising the tangential stress (see Table \ref{tab2} and Eqs.\,\eqref{nondim}, \eqref{lam}) we have the dimensional tangential stress amplitude scales as 
\begin{equation} \label{fintau2}  \tau_{12}^* \sim   \dfrac{\mu^* (d^*)^{1/2}}{(W^*\nu^*)^{1/2} (\lambda^*)^2  (1-a^2)^{3/4}} , 
\end{equation}
where $\lambda^*$ is the synovial fluid viscoelastic relaxation time, and  $\nu^*$ denotes the frequency of the joint motion. 
In particular Eqs.\, \eqref{ntscal},\, \eqref{fintau2}
give
$$ \frac{\tau_{12}^* } {\tau_{12}^{*,Newtonian}} \sim \dfrac{(d^*)^{3/2}}{(W^*)^{3/2} (\nu^*)^{1/2} (\lambda^*)^2  (1-a^2)^{3/4}} \in \frac 1{(1-a^2)^{3/4}}(1.1\times 10^{-10},2\times 10^{-4}),$$
where the range of values are determined from the upper and lower bounds from Table \ref{Tab:Pars}. This emphasises that elasticity introduces a multiple order of magnitude reduction in predicted tangential stress  across the physiological and pathophysiological parameter space. This prediction is a much more extensive reduction than indicated by shear thinning reductions in viscosity  measured for the synovial fluid. In particular,  synovial fluid viscosity at a frequency of 1Hz is reduced by an order of magnitude compared to very low frequency measurements (e.g. \cite{fam2007}, \cite{tirt1984} or Fig. 4 of \cite{liao2020}, based on \cite{schurz1987}),  generating the prediction of an order of magnitude drop in the tangential stress due to shear thinning  for the healthy joint at walking frequencies in the absence of synovial fluid elasticity.

The above  scaling observations have been independently confirmed in numerical simulations in Figs.~\ref{fig2}, \ref{fig3}, \ref{Fig:ODE}, \ref{surface}, though by the nature of numerical simulations the confirmation is for specific parameter values, while the analytical studies demonstrate the relations hold in much more generality. 
However, it should be noted that Fig.~\ref{Fig:ODE} highlights  fluctuations around this scaling are present as $\delta$ varies, with the same behaviour on varying the slip parameter $a$ and frequency $\nu$ also apparent in Eqs.~(\ref{scaling}),(\ref{fap1m}),(\ref{fap2m}) and, for $a$, in Fig.\,\ref{surface}. Nonetheless,  Fig.~\ref{Fig:ODE} does evidence that the scalings of Eqs.~(\ref{fintau1}),(\ref{fintau2})
certainly provides an accurate overall relation, and a simple upper bound for how the tangential stresses, and thus friction, varies with physical parameters for the oscillatory shearing of an elastic fluid.

Furthermore,  during the initial transient motion the amplitude of the non-dimensional and dimensional tangential stress scale as (Eqs.\,\eqref{nondim}, \eqref{trans}, \eqref{shst}) 
\begin{eqnarray*}  \tau^{{\small \mbox{trans}}}_{12} & \sim & \frac 1 {\sqrt{\beta}}= \frac 1 {\lambda \sqrt{1-a^2}}\sim   \frac{\lambda\nu^{1/2}(1-a^2)^{1/4}}{\delta^{1/2}}  \tau_{12}, \\ \tau^{*{\small \mbox{trans}}}_{12} &\sim &   \frac {\mu^*} {\lambda^* \sqrt{1-a^2}} \sim   \frac{\lambda^*(W^*\nu^*)^{1/2}(1-a^2)^{1/4}}{(d^*)^{1/2}}  \tau^*_{12}.
\end{eqnarray*}
While this transient tangential stress is much larger than the tangential stress after the decay of transients since $\delta\ll 1$, it is still much smaller than the tangential stress  associated with a Newtonian fluid, or an Oldroyd-B fluid, so that elasticity is also  predicted to be protective during motion initiation.

   More generally, the dependence of the slip parameter in these scalings and the interpretation of $a$ in terms of polymer  prolateness \citep{hinch2021} suggests that solutes of rod-like macromolecules with high persistence lengths may not be suitable for ameliorating friction. 
   {\bl Furthermore, as with other linear high molecular weight semi-flexible polymers, hyaluronan transitions from a random-coil to this rod-like conformation with decreasing molecular weight \citep{taweechat2020}.  The inflection point for the transition has been measured at $\sim$100-300~kDa \citep{weigel2017}. This lower molecular weight hyaluronan may be synthesised de-novo, or generated by either hyaluronidase mediated degradation or oxidative hydrolysis of the native macromolecule under pathological conditions, typically associated with inflammation and oxidative stress \citep{jiang2007,yang2012}. Specific to synovial fluid, mean molecular weights of 6-7~MDa are typically measured, with a lower molecular weight hyaluronan strongly correlated with pain and disease progression in osteoarthritis \citep{band2015}, and rheumatoid arthritis \citep{fam2007}. Thus $a$ is likely to increase with the pathological reduction in molecular weight of hyaluronan, providing another potential mechanism for increased friction, together with the related reduction in the elastic relaxation time, $\lambda^*$. 

   }


{\bl In Fig.\,\ref{surface} we also plot tangential stress predictions for the small amounts of elasticity associated with diseased joints. These show predictions of  dramatic rises in tangential stress. Nonetheless,  elasticity is still predicted to have the potential to be protective in pathophysiology,  even if the protection is lost for extremely small elasticity or extreme polymer prolateness. Furthermore, the impact of elasticity is fundamental and not captured by shear thinning models, as highlighted by considering steady state shear.  
However, elasticity also imparts  shear thinning with the effective viscosity in oscillatory shear   {\mg scaling with $\mu^*/(\nu^*)$ for an Oldroyd A or B fluid (\cref{obscal}), and scaling with  $\mu^*/(\nu^*)^{1/2}$ more generally (\cref{fintau2}).} This in turn indicates the prospect of   parameter identifiability difficulties in simultaneously extracting shear-thinning and elasticity parameters from experimental observations of synovial fluid to develop  rational rheological models in support of such directions. }



{\bl A further distinguishing feature of Newtonian and
{\mg Oldroyd A,\,B} fluids is that the tangential stress scales with the maximum velocity of the oscillatory relative motion. In contrast, the scaling relation of Eq.\, \eqref{fintau2} highlights that the tangential stress increases with reduced joint motion speed, $W^*$, once the  non-dimensional depth satisfies $\delta \ll 1,$ that is $ ~W^*T^*\gg d^*$, {\mg so that velocity scale has to be sufficiently large for the analysis to be valid. This constraint holds for joint motion and hence in the presence of elasticity,} tangential stresses within the joint, or more generally,  are predicted to reduce with increased tangential speeds across opposing surfaces.} 
Another distinguishing feature  is that Newtonian and {\mg Oldroyd A,\,B}  fluids do not exhibit rapid oscillation, but instead oscillations on the frequency of the joint movement (see e.g. Eqs.\,\eqref{ulconvect}, \eqref{ulconvect1}). 
{\mg In contrast, both Figs.\,\ref{fig2},\,\ref{fig3} and, for example, Eqs.\,\eqref{fap1m}, \eqref{fap2m},\,\eqref{freq} illustrate that once $$(1-a^2) \gg \delta^{2/3},$$   there are rapid  oscillations in the tangential stress with a frequency that scales with $1/\delta \gg 1 $ and thus with the reciprocal of the fluid thickness.}
{\bl The emergence of these oscillations, with  a completely different frequency scale compared to joint motion, is a fundamental prediction of this theory. While such high frequencies may be counter-intuitive, motion frequencies scaling inversely with drop volume have  been observed in viscoelastic drop dynamics \citep{sartori2022}, analogous to the frequency scaling with inverse depth here.}
Furthermore, Fig.~\ref{fig3} also indicates that the above scalings survive  to much deeper film thicknesses,
so that the above comments also apply much more generally to fluid lubrication, emphasising the fluid elasticity can reduce friction in more general oscillatory  settings rather than  solely joint mechanics.  
 
{\bl There are many possible generalisations of the work presented here.  This should include more complex constitutive relations, especially the impact of a  distribution of solute properties, as in practice the solute is not uniform within synovial fluid. Instead, hyaluronan exhibits a distribution of molecular weights. In turn, this  may be anticipated to induce a distribution of the elastic relaxation timescale and the slip parameter in the invariant derivative, resulting in  multiple viscoelastic timescales and responses within the same fluid. This further adds to the complexities of robustly estimating the rheological parameters from experiments and observations of synovial fluids. Despite such difficulties  
there  is  nonetheless a clear motivation for including elasticity within  the synovial fluid rheology of increasing sophisticated mechanical models of a joint that are under development, for example \citet{bridges2010,wu2017,liao2020} and \citet{hasnain2023}, not least because we have seen that  elasticity is predicted to fundamentally influence mechanics at the cartilage  interface.}

In summary, we have explored how the elastic properties of synovial fluid are predicted to  impact tangential stress  amelioration for  simple oscillatory shearing, with the derivation of scaling relations that summarise and bound how tangential stress depends on elastic parameters. These  scalings  are profoundly different from those of Newtonian fluids or  Oldroyd A,B fluids. In particular, the level of friction is predicted to be dramatically less with elasticity and the avoidance of extreme lubricating polymer prolateness. Furthermore, even though global fluid lubrication is not indicated in a joint, it is present as part of mixed mode lubrication. Hence,  the reductions of friction induced by elasticity might still be anticipated to be present, not least because refinements such as more complex geometries or constitutive relations would have to remove the predicted multiple orders of magnitude impact of elasticity on friction compared to Newtonian or Oldroyd A,B fluids. Thus further examination of synovial fluid elasticity is merited to characterise friction reduction in joints and its impact for increasingly sophisticated joint modelling.    In addition, the reductions in friction with fluid elasticity found in this study are also  valid on scales beyond that of physiological motion, more generally suggesting  relevance for lubrication and friction amelioration for the oscillatory motion of  opposing surfaces.

\section*{Appendix. Simple Harmonic Approximation via Stationary Phase}

We consider simplifications for the tangential stress when the relative motion of the boundaries is given by a simple harmonic driving 
$$ U(t) = \sin(2 \pi\nu t) , $$
of unit maximum magnitude, by using the stationary phase approximation given by 
Eq.~\eqref{stphap}. Initial transients are neglected. Fixing the time $t$, we have that the phase is stationary when $t=t_*$ with $U(t_*)=0,$ so that 
$$ t_*= \dfrac{n}{2\nu},  ~~~~ U'(t_*) = 2\pi\nu (-1)^n, ~~~ |U'(t_*)| = 2\pi\nu , ~~~ \mbox{sgn} (U'(t_*)) = (-1)^n,  $$
where $n$ is natural and $0\leq t_*\leq t$. We also let $N$ denote the maximal $n$ for a given $t$, so that $t^{max}_* = N\pi/[2\nu]$ and noting that, since $U(0)=0$,  $N$ always exists. Then from 
Eq.\eqref{stphap} we have 
\begin{eqnarray}  \nonumber  \tau_{12} &\approx& 
\dfrac{\delta^{1/2} \e^{-(t-t_*^{max})/\lambda}}{\lambda^{1/2}\nu^{1/2}\beta^{3/4}} 
\sum_{n=0}^N \e^{-(t_*^{max}-t_*)/\lambda} 
\mbox{Im} \left(\exp\left[ -\mathrm{i}\dfrac{\pi}{4}(-1)^n+\dfrac{\mathrm{i}\sqrt{\beta}}{\delta\lambda} \int^t_{t_*}  \mathrm{d}r \, U(r) \right] 
\right).   
\end{eqnarray} 
We take $N$ to be even below  in the first instance. With 
$$ \int^t_{t_*}  \mathrm{d}r \, U(r) = \dfrac 1 {2\pi\nu}\left[(-1)^n-\cos(2\pi\nu t)\right],$$
and the definition 
$$ \xi :=  -\dfrac \pi 4 + \dfrac{\beta^{1/2}}{2\pi \nu \delta \lambda}, $$
we have 
$$ \exp\left[ \mathrm{i}\dfrac{\pi}{4}(-1)^n+\dfrac{\mathrm{i}\sqrt{\beta}}{\delta\lambda} \int^t_{t_*}  \mathrm{d}q \, U(q) \right]  = \exp\left[\mathrm{i}\xi(-1)^n -\dfrac{\mathrm{i} \beta^{1/2}}{2\pi \nu \delta \lambda} \cos(2\pi\nu t) \right].
$$
 Thus in the approximation to $\tau_{12}$ we need the summation
$$ \sum_{n=0}^N \exp\left[-\left(\frac{N-n}{2\nu\lambda}\right) + \mathrm{i}\xi(-1)^n\right] = \sum_{n=0}^N \exp\left[-\left(\frac{n}{2\nu\lambda}\right) + \mathrm{i}\xi(-1)^n\right], $$
where the second term arises from reversing the order of the summation and recalling  that we have taken $N$ even in the first instance. Thus the summation reduces to 
$$  \e^{\mathrm{i}\xi} \sum_{m=0}^{N/2} \left(\exp\left[-\left(\frac{1}{\nu\lambda}\right)\right]\right)^m + 
\e^{-\mathrm{i}\xi} \exp\left[-\left(\frac{1}{2\nu\lambda}\right)\right]\sum_{m=0}^{-1+N/2} \left(\exp\left[-\left(\frac{1}{\nu\lambda}\right)\right]\right)^m 
.$$
For $t$ sufficiently large, so that $N=2\nu t_*^{max}$ is sufficiently large to allow the neglect of  exponentially small corrections,   the summation reduces to 
$$ \dfrac {  \e^{\mathrm{i}\xi} + \e^{-\mathrm{i}\xi} \e^{-1/(2\nu\lambda)} }  {1-\e^{-1/(\nu\lambda)} }   .$$ 
Thus we have 
\begin{eqnarray}  \nonumber  \tau_{12} &\approx& 
\dfrac{\delta^{1/2} \e^{-(t-t_*^{max})/\lambda}}{\lambda^{1/2}\nu^{1/2}\beta^{3/4}} \mbox{Im} \left( \exp\left[ -\dfrac{\mathrm{i} \beta^{1/2}\cos(2\pi\nu t)}{2\pi \nu \delta \lambda}  \right] \dfrac {  \e^{\mathrm{i}\xi} + \e^{-\mathrm{i}\xi} \e^{-1/(2\nu\lambda)} }  {1-\e^{-1/(\nu\lambda)} } \right)  .
\end{eqnarray}
Recalling $\beta = \lambda^2(1-a^2)$ and with the definitions
\begin{eqnarray*}
s_1 &:=& \sin   \left( \dfrac{\beta^{1/2}}{2\pi\nu\lambda\delta}(1-\cos(2\pi\nu t))-\dfrac \pi 4  \right) = \sin   \left( \dfrac{(1-a^2)^{1/2}}{2\pi\nu\delta}(1-\cos(2\pi\nu t))-\dfrac \pi 4  \right) \\
s_2 &:=&  \sin\left( \dfrac{\beta^{1/2}}{2\pi\nu\lambda\delta}(1+\cos(2\pi\nu t))-\dfrac \pi 4  \right)=\sin\left( \dfrac{(1-a^2)^{1/2}}{2\pi\nu\delta}(1+\cos(2\pi\nu t))-\dfrac \pi 4  \right)
\end{eqnarray*}
this reduces to 
\begin{eqnarray}\label{fap1}
\tau_{12} &\approx& 
\dfrac{\delta^{1/2} \e^{-(t-t_*^{max})/\lambda}} {\lambda^{2}\nu^{1/2}(1-a^2)^{3/4}(1-\e^{-1/(\nu\lambda)})} \left(s_1 - \e^{-1/(2\nu\lambda)}s_2\right) . 
\end{eqnarray}
Recall the above is for $N$ even;  an analogous approximation for $N$ odd instead generates
\begin{eqnarray}\label{fap2}
\tau_{12} &\approx& 
\dfrac{\delta^{1/2} \e^{-(t-t_*^{max})/\lambda}} {\lambda^{2}\nu^{1/2}(1-a^2)^{3/4}(1-\e^{-1/(\nu\lambda)})} \left(  \e^{-1/(2\nu\lambda)}s_1-s_2\right) .
\end{eqnarray}

\section*{Rights Retention Statement}
For the purpose of Open Access,  a CC-BY public copyright licence is applied to any Author Accepted Manuscript version arising from this submission.

\section*{Data Accessibility}
The data in this paper is available from the University of Oxford research archive at 

http://dx.doi.org/xxx/xxx.

\noindent 
{\it Peer review note.}  A University of Oxford research archive doi has not been generated for peer review, as this requires final publication metadata. For the purpose of peer review, the data in this paper may be accessed by editors and reviewers from the link: 

\href{https://www.dropbox.com/scl/fi/dauyuzvsgtjmd882d7qy8/DatafilesFinal.zip?rlkey=3ed4xiq00zoea2dgcw5rkbwcl&dl=0}{DataFilesLinkForPeerReview}

\bibliographystyle{jfm.bst}
\bibliography{library.bib}
\end{document}